\documentclass[lettersize,journal]{IEEEtran}

\usepackage{longtable}
\usepackage[printonlyused]{acronym}

\usepackage[dvipsnames]{xcolor}
\usepackage{wrapfig}
\usepackage{caption}
\usepackage{textcomp}

\usepackage{graphicx}
\usepackage{amsmath,amssymb,amsfonts}
\usepackage{algorithmic}

\usepackage{subcaption}
\usepackage[inline]{enumitem}
\acrodef{3D}{three-dimensional}
\acrodef{3GPP}{3rd generation partnership project}
\acrodef{5G}{fifth generation}
\acrodef{6G}{sixth generation}
\acrodef{A2C}{actor-critic}
\acrodef{ABG}{alpha-beta-gamma}
\acrodef{AoD}{angle of departure}
\acrodef{AoA}{angle of arrival}
\acrodef{ANN}{artificial neural networks}
\acrodef{AUC}{area under curve}
\acrodef{AWGN}{additive white gaussian channel}
\acrodef{BB}{broad beam}
\acrodef{BS} {base station}
\acrodef{BSs} {base stations}
\acrodef{BTS}{bayesian-thompson sampling}
\acrodef{CAB}{contextual multi-armed bandit}
\acrodef{CS}{compressive sensing}
\acrodef{CNN}{convolution neural network}
\acrodef{CSI}{channel state information}
\acrodef{DFT}{discrete Fourier transform}
\acrodef{DNN}{deep neural networks}
\acrodef{DRL}{deep reinforcement learning}
\acrodef{EIRP}{effective isotropic radiation power}
\acrodef{hDRL}{hierarchical deep reinforcement learning}
\acrodef{DQN}{deep Q-network}
\acrodef{hDQN}{hierarchical deep Q-network}
\acrodef{ESN}{echo state network}
\acrodef{FM}{frequency modulated}
\acrodef{FML}{fast machine learning}
\acrodef{FPR}{false positive rate}
\acrodef{fsp}{free space path loss}
\acrodef{gNB}{5G NR base station}
\acrodef{gUCB}{greedy upper confidence bound}
\acrodef{GLOBECOM}{Global Communication Conference}
\acrodef{GPS}{global positioning system}
\acrodef{HAR}{human activity recognition}
\acrodef{HuMRaBD}{\underline{Hu}man \underline{M}illimeter wave \underline{Ra}dio \underline{B}locakge \underline{D}etection}
\acrodef{IA}{initial access}
\acrodef{ICONICAL}{IIoT CONnectivity for mechanICAL systems}
\acrodef{IID}{independent and identical distribution}
\acrodef{IR}{Infrared}
\acrodef{LO}{local oscillator}
\acrodef{LoS}{line-of-sight}
\acrodef{MAB}{multi-armed bandit}
\acrodef{MIMO}{multiple input multiple output}
\acrodef{ML}{machine learning}
\acrodef{MLP}{multi layer perceptron}
\acrodef{MSE}{mean squared error}
\acrodef{OTA}{over-the-air}
\acrodef{NB}{narrow beam}
\acrodef{NIC}{network interface card}
\acrodef{nLoS}{non-line-of-sight}
\acrodef{NN}{neural networks}
\acrodef{NR}{new radio}
\acrodef{OFDM}{orthogonal frequency division multiple access}
\acrodef{mmWave}{Millimeter wave}
\acrodef{MDP}{markov decision process}
\acrodef{MLP}{multi layer perceptron}
\acrodef{POMDP}{partially observable Markov decision process}
\acrodef{PWR}{passive WiFi radar}
\acrodef{RBD}{receiver beam direction}
\acrodef{ReLU}{rectifier linear units}
\acrodef{RL}{reinforcement learning}
\acrodef{RF}{radio frequency}
\acrodef{RIM}{RSSI Information Matrix}
\acrodef{ROC}{receiver operating characteristic}
\acrodef{RSS}{received signal strength}
\acrodef{RSSI}{received signal strength indicator}
\acrodef{SDR}{software defined radio}
\acrodef{SGD}{stocastic gradient descent}
\acrodef{SNR}{signal-to-noise ratio}
\acrodef{SSB}{synchronisation signal block}
\acrodef{SS}{synchronisation signal}
\acrodef{SCS}{sub-carrier spacing}
\acrodef{TPR}{true positive rate}
\acrodef{TTU}{travel time unit}
\acrodef{TDMA}{time domain multiple access}
\acrodef{UAV}{unmanned aerial vehicle}
\acrodef{UAVs}{unmanned aerial vehicles}
\acrodef{UE}{user equipment}
\acrodef{UEs}{user equipments}
\acrodef{ULA}{uniform linear array}
\acrodef{UMa}{urban macro-cellular}
\acrodef{UPA}{uniform planar array}
\acrodef{USRP}{universal software radio peripheral}
\acrodef{UWB}{Ultra-wide band}
\acrodef{V2X}{vehicle-to-everything}
\acrodef{QoS}{quality of service}
\acrodef{WiFi}{wireless fidelity}

\begin{document}


\title{Non-contact Multimodal Indoor Human Monitoring Systems: A Survey}

\author{
\IEEEauthorblockN{Le Ngu Nguyen$^{\star}$ \enspace Praneeth~Susarla$^{\star}$ \enspace Anirban Mukherjee$^{\star \dagger}$ \enspace Manuel Lage Cañellas$^{\star}$ \\ Constantino Álvarez Casado$^{\star}$ \enspace Xiaoting~Wu$^{\star}$ \\ Olli~Silvén$^{\dagger}$ \enspace Dinesh Babu Jayagopi$^{\dagger}$ \enspace Miguel Bordallo López$^{\star}$}

\IEEEauthorblockA{$^{\star}$ Center for Machine Vision and Signal Analysis, University of Oulu, Finland \\
$^{\dagger}$ International Institute of Information Technology, Bangalore, India}

\thanks{Praneeth~Susarla, Le Ngu Nguyen, Anirban~Mukherjee, and Manuel~Lage~Cañellas have contributed equally to this work.}
\thanks{This research has been supported by the Academy of Finland 6G Flagship program under Grant 346208 and PROFI5 HiDyn under Grant 32629, and the InSecTT project, which is funded under the European ECSEL Joint Undertaking (JU) program under grant agreement No 876038.}
}


\maketitle


\begin{abstract}
Indoor human monitoring systems leverage a wide range of sensors, including cameras, radio devices, and inertial measurement units, to collect extensive data from users and the environment. These sensors contribute diverse data modalities, such as video feeds from cameras, received signal strength indicators and channel state information from WiFi devices, and three-axis acceleration data from inertial measurement units.
In this context, we present a comprehensive survey of multimodal approaches for indoor human monitoring systems, with a specific focus on their relevance in elderly care.
Our survey primarily highlights non-contact technologies, particularly cameras and radio devices, as key components in the development of indoor human monitoring systems.
Throughout this article, we explore well-established techniques for extracting features from multimodal data sources.
Our exploration extends to methodologies for fusing these features and harnessing multiple modalities to improve the accuracy and robustness of machine learning models.
Furthermore, we conduct comparative analysis across different data modalities in diverse human monitoring tasks and undertake a comprehensive examination of existing multimodal datasets.
This extensive survey not only highlights the significance of indoor human monitoring systems but also emphasizes their versatile applications.
In particular, we emphasize their critical role in enhancing the quality of elderly care, offering valuable insights into the development of non-contact monitoring solutions applicable to the needs of aging populations.
\end{abstract}

\section{Introduction}
\label{sec:introduction}

\begin{figure*}
\includegraphics[width=\textwidth]{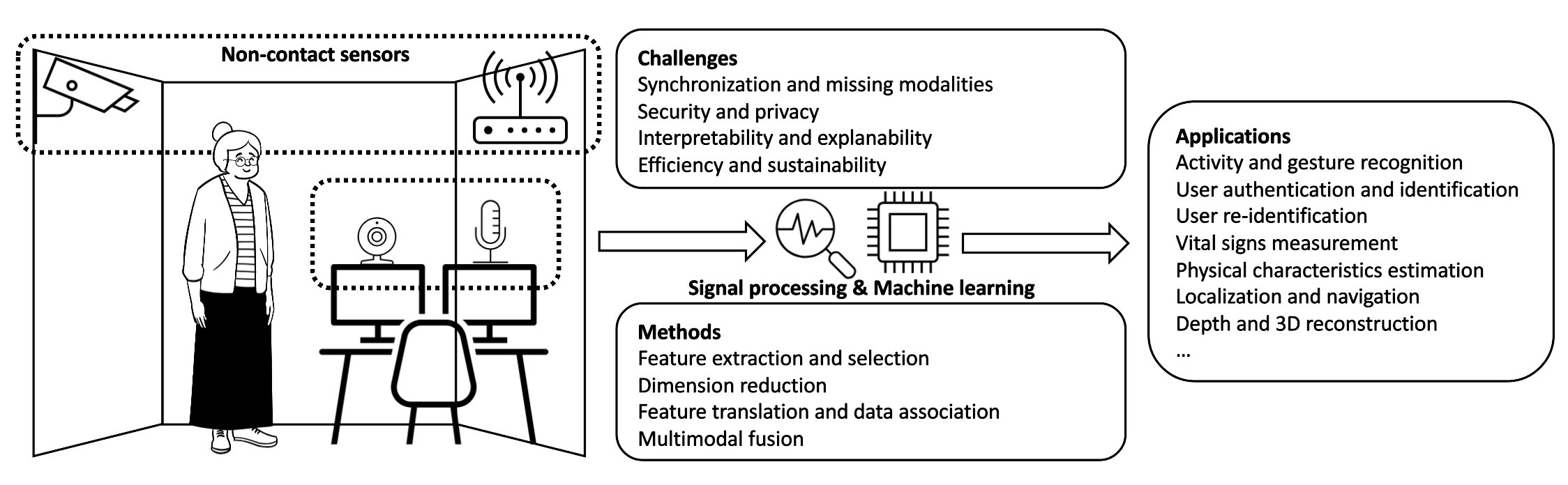}
\caption{Non-contact elderly monitoring system}
\label{fig:teaser}
\end{figure*}

One important objective of the \textit{WHO Global strategy and action plan on ageing and health} is to customize healthcare systems to the needs of elderly people~\cite{world2023progress}.
Indoor human monitoring is an essential component of an elderly care system, providing critical decision-making, personalized care, emotional support or interdisciplinary collaboration, ultimately enhancing the overall effectiveness and reliability of remote care delivery~\cite{Gokalp2013}.
Indoor human monitoring systems observe the indoor environments using sensors such as cameras~\cite{Aggarwal2014}, WiFi devices and radars~\cite{Fernandes2022}, or wearable devices~\cite{wang2019wearablesensors}. Every sensor possesses its own set of advantages and disadvantages, making them unique in their capabilities and limitations. For example, a camera can capture more reliable information about the users but can also intrude on their privacy leading to discomfort and objections from the individuals being monitored. Ambient WiFi signals can be analyzed to infer human activities in a non-intrusive manner with limited reliability. Radar devices can measure subtle chest displacements to estimate vital signs but may struggle to identify objects and are expensive to monitor objects with optimal performance. While the unimodal usage of these sensors posses limitations, a combination of visual data and wireless signals is more robust and practical, providing holistic information to implement non-contact human monitoring systems~\cite{Haque2020}. 

Multimodal systems provides a multi-aspect view on current situations.
For instance, in robotics and smart factories, multiple sensors attached to different locations and devices facilitate contextual understanding and intelligent decision-making functions~\cite{Li2021_Crowd}.
A multimodal system can adapt itself to different scenarios based on the user's settings, providing a flexible approach that respects individual requirements and privacy concerns.
For example, while cameras and radio devices could be acceptable in a living room, only radio equipment could be installed in a bedroom in order to keep the user's privacy. This adaptability extends to how different modalities collaborate. To serve specific tasks optimally, such as activity recognition in indoor monitoring scenarios, modalities can be paired in innovative ways. A camera, for example, can guide a radar device to the best location for measuring respiration and heart rate~\cite{shokouhmand2022camera}. Similarly, radio-based positioning data can be utilized as input for a vision-based human detection algorithm~\cite{Qiu2022detection}. Through these combined approaches, a multimodal system enhances accuracy, reliability, and context awareness~\cite{Li2021_Crowd}. It also offers the benefits of being robust, redundant, and adaptable to various monitoring needs while preserving user privacy~\cite{shokouhmand2022camera}.

In this work, we survey the existing multimodal systems for different human monitoring tasks in indoor scenarios. Indoor human monitoring systems typically involve various tasks such as activity recognition, gait analysis, vital signs measurement, user localization, and user identification.
Elderly care, particularly in the context of telecare systems~\cite{Gokalp2013}, has gained immense consideration due to the increasing number of senior individuals living independently~\cite{Plothner2019}~\cite{Stanford2002}.
Our research primarily focuses on this domain, while also shedding light on other applications such as assisting patients at home.
These human monitoring systems not only bolster a sense of safety but also detect crucial events (e.g., falling and onset of diseases) in order to facilitate timely intervention.
Over time, the continuous monitoring offered by these systems aids healthcare professionals in observing the progression of chronic conditions, laying the foundation for tailored treatment plans. In order to gain insights on how to realize a state-of-the-art elderly monitoring system, we explore a wide range of sensors and machine learning techniques, with a special focus on non-contact methods.
We followed the Preferred Reporting Items for Systematic Reviews and Meta-Analyses (PRISMA) guideline~\cite{Page2021} to search and select relevant studies for this survey.
First, we defined the keywords for searching on Google Scholar, IEEE Xplore, ACM Digital Library, and Springer Link.
The relevent studies should include these keywords: \begin{enumerate*}
    \item \textit{human monitoring},
    \item \textit{indoor environments}, \textit{indoor settings}, \textit{indoor scenarios},
    \item \textit{multimodal},
    \item \textit{multiple sensors}, \textit{multi-sensor}, and
    \item \textit{non contact}.
\end{enumerate*}
Then, we refined the found articles using the inclusion and exclusion criteria:
\begin{itemize}
    \item Inclusion criteria: \begin{enumerate*}
        \item Articles are peer-reviewed.
        \item Articles are related to human monitoring.
        \item Articles utilize multiple sensors to collect and process multimodal data to implement human monitoring applications, as well as showing the advantages of multi-sensor systems.
        \item Techniques in the articles can be applied to elderly monitoring systems.
        \end{enumerate*}
    \item Exclusion criteria: \begin{enumerate*}
        \item Articles focus on outdoor applications of multi-sensor systems and multimodal data (such as autonomous driving).
        \item Articles utilize multiple sensors that are carried by the users or attached to the human bodies.
        \item Articles analyze multimodal data for applications not directly related to human subjects, such as vehicle or animal tracking.
        \end{enumerate*}
\end{itemize}

The earliest related work that we can find is published by Krahnstoever~\textit{et al.}~\cite{krahnstoever2005activity} in 2005, which embedded Radio Frequency Identification (RFID) tags into objects in order to augment visual data for object tracking and activity recognition.
After collecting the relevant articles, we analyzed them to identify tasks required in a multi-sensor system for monitoring elderly people. 
These comprehensive monitoring systems undertake a series of crucial tasks to ensure the well-being and safety of elderly residents~\cite{Stanford2002}.
Initially, they focus on the precise recognition of elderly users and their daily activities, forming a baseline understanding of their routines.
Subsequently, these systems take on essential responsibilities such as conducting gait analysis and monitoring vital signs to track evolving health conditions, offering timely assistance as needed. In addition to health-related functions, these systems extend their capabilities to encompass surveillance tasks, including the creation of 3D models of indoor environments for precise localization of elderly individuals during suspicious movements, thus ensuring not only their physical security but also their overall well-being in the context of independent living. An example of an advanced non-contact human monitoring system like elderly care can be summarized in terms of sensors, methods, challenges, and applications as shown in Figure~\ref{fig:teaser}.
We organize the survey around the techniques to implement these tasks, which are essential in the context of elderly care and applicable to related human monitoring systems.
The tasks include activity recognition (see Section~\ref{sec:activityrecognition}), vital sign measurement (see Section~\ref{sec:vitalsignmeasurement}), user identification (see Section~\ref{sec:identification}), localization (see Section~\ref{sec:localization}), and 3D modeling of the environment (see Section~\ref{sec:3d}).
In addition, we complement these sections with a small set of articles related to alternative sensing methods, which give context to the survey.

\begin{figure*}[ht!]
 \begin{center}
   \includegraphics*[width=0.99\textwidth]{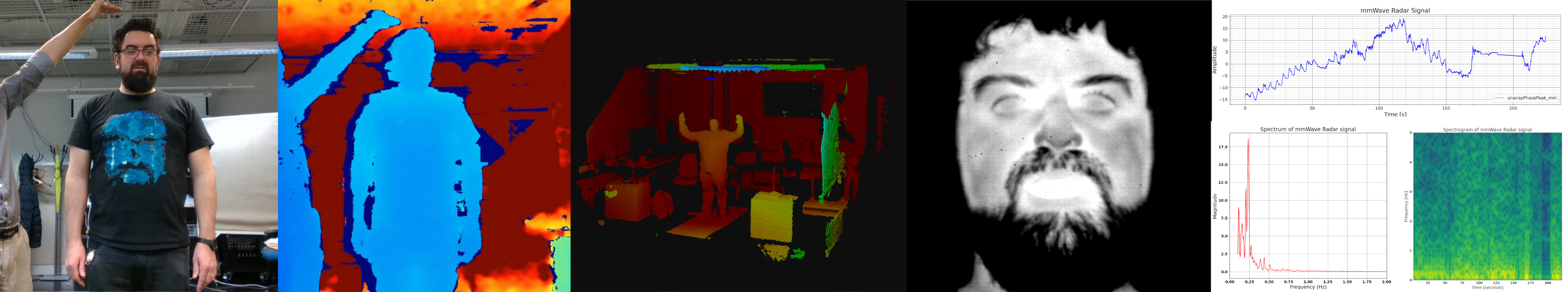}
 \end{center}
 \vspace{-3mm}
 \caption{From left to right: RGB camera, Depth camera, 3D Depth camera, Thermal camera and mmWave radar data.}
 \label{fig:modalities_sensors}
 \vspace{-3mm}
\end{figure*}

For developing non-contact human monitoring systems, we focus on the usage of non-contact sensors such as cameras and radio devices.
We also include their interactions with other complementary sensors such as on-body inertial measurement units, pressure plates, microphones, and wearable vital sign monitoring devices.
These main sensors, each with its unique strengths and limitations, form the core of elderly monitoring solutions.
They offer complementary insights and capabilities that, when combined thoughtfully, create a comprehensive monitoring solution tailored to specific needs of the users and constraints of the environment.
An example of multimodal data captured with a diverse set of sensors is depicted in Figure \ref{fig:modalities_sensors}.

\textbf{RGB Cameras}: Red-Green-Blue (RGB) cameras are the visual workhorses of elderly monitoring systems. They excel at capturing rich visual information, including color, texture, and shape, providing a comprehensive view of the monitored environment~\cite{Sun2022_survey}. Their versatility is evident in numerous applications, from tracking user activities to detecting anomalies. However, they are susceptible to environmental factors such as varying light conditions and occlusions, which can impact their performance. Additionally, privacy concerns often arise due to the intrusive nature of constant visual monitoring. Nevertheless, their ubiquity and ability to offer real-time insights make them indispensable in many monitoring scenarios.

\textbf{Depth Cameras}: Depth cameras, utilizing either time-of-flight or stereoscopic methods, are specialized sensors designed to capture accurate 3D spatial and shape information. These characteristics make them particularly well-suited for tasks that demand precise depth perception, such as fall detection and gesture recognition. One drawback is that they do not capture color or texture data, features that are sometimes essential for specific applications. To overcome this limitation, depth cameras are often integrated with RGB cameras to form RGB-D systems \cite{Jianwei2023RGBDReview}. These hybrid sensors offer a balanced data set that includes color, texture, and depth, thereby contributing to a more comprehensive monitoring solution.

\textbf{Thermal Cameras}: Infrared cameras are effective in dark environments, making them valuable for monitoring when visibility is low. However, they also lack color and texture information and are sensitive to heat sources. Their use is ideal for scenarios where visual information is not critical, and the focus is on temperature-related monitoring. Their cost limits their usage to certain scenarios. Consequently, there are few public datasets of thermal data and models trained on these datasets, comparing to those of RGB and RGB-D modalities.

\textbf{Radio Devices}: Radio-frequency sensors, including radar and WiFi devices, have gained prominence for their ability to operate in non-line-of-sight scenarios and their robustness in variable lighting conditions~\cite{yousefi2017survey}. These sensors have applications in tracking user movements, detecting vital signs, and even through-wall sensing. Nevertheless, they share a common limitation with other non-visual sensors, as they lack color, texture, and shape data. They often require sophisticated postprocessing techniques to extract interpretable information, but their unique capabilities make them valuable components in multimodal elderly monitoring systems.

\textbf{Complementary Sensors}: These sensors collectively form a group that complements the strengths and weaknesses of visual and radio sensors in elderly monitoring systems. Inertial measurement units, commonly found in devices like smartphones and smartwatches, offer cost-effective means of tracking user movements and activities but lack visual information~\cite{wang2019wearablesensors}. Ambient sensors~\cite{Ciliberto2021}, which measure environmental factors like temperature, humidity, and object usage, provide insights into the living conditions but offer limited information about individual users. Pressure-sensitive sensors integrated in surfaces, such as the floor, are versatile, aiding in gait analysis and fall detection; however, their installation and maintenance can be complex. Audio sensors, represented by microphones, have numerous applications and are commonly integrated into devices, making them a practical choice for audio-related monitoring~\cite{Morita2023}. Nevertheless, they do not provide visual information, and privacy concerns regarding voice data are significant. These complementary sensors can enhance the overall monitoring system by providing context and supplementary data, making them valuable additions to not only an elderly monitoring system but also to general human monitoring solutions.

In our survey, we study the utilization of multimodal approaches for different tasks and scenarios in indoor non-contact human monitoring systems, focusing on monitoring elderly individuals~\cite{Stanford2002}.
This survey is useful for researchers aiming to leverage multimodal approaches in developing an elderly monitoring solution and related systems.
We aims to link the sensors and methods to provide a logical view of the multimodal approaches.
Besides offering a deep dive into current practices, it identifies gaps in the existing literature and suggests prospective avenues for exploration, especially in the context of multi-modality.

\noindent\fbox{\begin{minipage}{0.96\columnwidth}
    The contribution of this survey is listed as follows:
    \begin{itemize}
        \item We investigate systems that utilize vision and radio-frequency signals, with the supplement of other sensors.
        \item We summarize methods of combining vision, radio, and other sensing modalities in different levels: early, intermediate, and late fusion (or decision fusion).
        \item We highlight the advantages of integrated vision and radio devices, in comparison to other systems that leverage only cameras or radio sensing.
        \item We present challenges and limits of the existing multimodal human monitoring systems, as well as proposing potential solutions to handle these issues.
    \end{itemize}
\end{minipage}}

The existing surveys on indoor human monitoring systems concentrate on either individual modalities or one task, such as vision-based activity recognition~\cite{Aggarwal2014}~\cite{yuan2022survey}, RGB-based and skeloton-based activity recognition~\cite{Wang2023}, action recognition based on wearable sensors~\cite{wang2019wearablesensors}, RF-based human behaviour recognition~\cite{yousefi2017survey}~\cite{Fernandes2022}~\cite{wang2015review}, and vital signs measurement based on WiFi CSI~\cite{Soto2022}.
Among surveys focusing on multimodal approaches, we can found those investigating activity recognition~\cite{Sun2022_survey}, object detection ~\cite{tang2022survey}, fusion methods for multimodal data~\cite{Baltruvsaitis2018Survey}, and classification models~\cite{Sleeman2022}.
In these surveys, multimodal approaches have shown their advantages.
For example, Fan~\textit{et al.}~\cite{Fan2020} utilized a video-based captioning method to improve the radio-based approach in diary generation.
Their combination leveraged an accurate modality to enhance a privacy-aware sensing device.
In vital sign monitoring, a camera can guide the radar to the sensing region on the human body~\cite{shokouhmand2022camera}.
These examples have proved that an optimal integration of multiple modalities can improve and extend the utility of human sensing systems while mitigating the risk involving the collection of sensitive data.

\section{Human monitoring tasks on multimodal data}
In this section, we identify important tasks that are required in a elderly monitoring system.
They are: activity recognition, vital sign measurement, identification, localization, and 3D scene modeling.
We describe popular techniques that are utilized to implement these tasks.


\subsection{Activity recognition}
\label{sec:activityrecognition}

In this section, we investigate human activity recognition systems using cameras, microphones, wearable sensors, and radio devices.
The input of an activity recognition model is a feature vector $v \in \mathbb{R}^k$, where $k$ is the number of features, and its output is an activity such as standing, walking, falling, and lying.
The feature vectors are extracted from one sensor or combined from multiple modalities.
Sun~\textit{et al.}~\cite{Sun2022_survey} provided a comprehensive survey on human activity recognition from single-modality to multi-modality perspectives.

\begin{table*}
\caption{Human activity recognition using multi-modal non-contact human assisting systems}\label{tab:activity_recognition}

\begin{tabular}{|p{3cm}|p{3cm}|p{4cm}|p{6cm}|}
\hline
\textbf{Articles}  & \textbf{Sensors}    & \textbf{Multimodal approaches} & \textbf{Objectives} \\ \hline
Zhao~\textit{et al.}~\cite{zhao2018through}&   sub-GHz WiFi, RGB camera &   radio features with camera supervision & Human pose estimation in indoor activities such as walking, sitting, taking stairs, waiting for elevators, opening doors, and talking to friends \\ \hline

Zhao~\textit{et al.}~\cite{Zhao2021} & Force sensor, Kinect camera, and wearable accelerometer & Correlative memory neural network (CorrMNN) to measure the correlation in bimodal gait data & Learning gait differences between three \underline{neurodegenerative diseases}, between patients with different severity levels of Parkinson's disease, and between healthy individuals and patients \\ \hline

Shao~\textit{et al.}~\cite{Shao2022} & 4 Kinect cameras (skeleton data and RGB image data) & Long Short-Term Memory (LSTM) model for skeleton data, two Convolutional Neural Network (CNN) models for RGB videos (front and side perspectives), fusing silhouette features, and decision level fusion & Detection \underline{depression} in a dataset of 200 postgraduate students (86 depressive ones) \\ \hline

%
%
%
%
%
Zou~\textit{et al.}~\cite{zou2019wifi}    & sub-GHz WiFi, RGB camera   &   decision-level & sitting, standing, walking when the light is on or off \\ \hline
%
%
Ardianto~\textit{et al.}~\cite{ardianto2018infrared} & Infrared (IR), RGB-D  &   feature fusion & standing, walking, and running \\ \hline
De~\textit{et al.}~\cite{de2020infrared}   & Infrared, RGB-D  &   feature fusion & NTU RGB+D dataset: 82 daily individual actions, 26 two-person actions, and 12 \underline{medical condition actions} (neeze/cough, staggering, falling down, headache, chest pain, back pain, neck pain, nausea/vomiting, fan self, yawn, stretch oneself, blow nose) \\ \hline
Krahnstoever~\textit{et al.}~\cite{krahnstoever2005activity}   &  RFID in objects, RGB camera  &   feature fusion & recognizing human-object interaction: person approaching (an object), taking/examining/reading/replacing an object, and person leaving \\ \hline
%
%
%
Memmesheimer~\textit{et al.}~\cite{memmesheimer2020} & WiFi, Camera, inertial sensor, motion capture &  transforming all data to one common representation (2D images) as input for CNN & Daily activities from NTU RGB+D, UTD-MHAD (camera and IMU), and Simulate (RGBD) \\ \hline
Li~\textit{et al.}~\cite{Li2018} & UWB radar, Wearable inertial sensors &  & Activity recognition and \underline{fall} detection \\ \hline


Fan~\textit{et al.}~\cite{Fan2020} & Camera, WiFi & Multimodal feature alignment & Daily activity recognition and diary, using video-based captioning to improve the performance of the radio-based method. \\ \hline

\end{tabular}
\end{table*}

Under the unimodal approaches, recent \ac{HAR} systems have proposed training the learning models from a single data source information such as video, audio, acceleration, and wireless signal. 
Unimodal activity recognition using vision or radio waves have been extensively studied in the literature~\cite{robertson2006general}\cite{nie2021pose2room}\cite{han2005human}. RGB videos were leveraged by Robertson~\textit{et al.}~\cite{robertson2006general} to perform activity recognition. Nie~\textit{et al.}~\cite{nie2021pose2room} analyzed the \ac{3D} scenes by observing human trajectory in the environment. Han~\textit{et al.}~\cite{han2005human} used thermal infrared imaging data and performed human motion detection independent of the lighting conditions, colors of human surfaces, or backgrounds. Using radio waves, Li~\textit{et al.}~\cite{li2019making} performed action recognition in smart home systems where cameras are difficult to utilize due to inadequate lighting, and privacy concerns.

Multi-modal \ac{HAR} applications utilize the data from multiple modalities such as RGB, depth, infrared camera, wearable sensors, and skeleton data. Most of the recently proposed multi-modal \ac{HAR} applications are applied on the MSRDailyActivity3D, UTH-MHAD, NTU RGB+D benchmark multimodal datasets and are broadly classified into co-learning-based and fusion-based methods. Co-learning methods explore the learned knowledge from auxiliary sensor data and use them to assist the learning processor of a different multi-modal sensor data source. Fusion-based methods generally involve multiple combinations of visual modalities such as RGB - depth - skeleton data, visual and non-visual modalities such as audio - acceleration - RGB - depth.

Recent works~\cite{bocus2022uwbdataset, bocus2022operanet, guo2018, Li2019} proposed multimodal datasets for \ac{HAR} using the modalities from different radio, radar and camera information. Bocus~\textit{et al.}~\cite{bocus2022operanet} proposed a \ac{HAR} multimodal dataset using Wi-Fi \ac{NIC} interface, \ac{PWR}, \ac{UWB}, vision camera and infrared sensor information. Gou~\textit{et al.}~\cite{guo2018} proposed a human gesture-based activity recognition by combining Kinect-camera information with CSI information of the WiFi devices. The proposed work also published a WiAR dataset. Using the OPERAnet dataset~\cite{bocus2022operanet}, Koupai~\textit{et al.}~\cite{koupai2022activity} implemented a fusion-based vision transformer approach for HAR on PWR radar, CSI data, and vision data. Similarly, Li~\textit{et al.}~\cite{Li2019} combined camera with FMCW radar information for activity recognition in dark and occlusion conditions. The authors also released the multimodal dataset.
Although the aforementioned multi-modality fusion-based HAR methods have achieved promising results on some benchmark datasets, the task of effective modality fusion is still largely open. Specifically, most of the existing multi-modality methods have complicated architectures that require high computational costs.
In Table~\ref{tab:activity_recognition}, most work focused on daily activities of healthy subjects.
Some of them aim to detect falling and health-related movements (e.g., sneezing, coughing, and staggering)~\cite{de2020infrared}~\cite{Li2018}.
In an elderly monitoring system, it is useful to integrate an automatic diary generator~\cite{Fan2020}, which may help medical experts to assess long-term conditions and treatment effectiveness.

Analysing the walking parameters of a patient provides useful information about movement disorders and neurodegenerative diseases.
It is challenging to distinguish between normal and abnormal gait features as well as between different severity levels of diseases.
To overcome this problem, Zhao~\textit{et al.}~\cite{Zhang2021} aggregated data from multiple sensors to learn the differences between three neurodegenerative diseases, between patients with different severity levels of Parkinson's disease, and between healthy individuals and patients.
Although the subjects in~\cite{Shao2022} were not elderly people, Shao~\textit{et al.}~\cite{Shao2022} illustrated the feasibility of detecting depression with Kinect cameras.
For RGB data, they combined silhouette features from the front and the side view to produce predictions, which were fused at the decision level with the skeleton-based LSTM outputs.
These methods are useful for an elderly monitoring system, especially for long-term assessment of neurodegenerative diseases and mental disorders.

Human Activity Recognition (HAR) remains an important area of research in the domain of elderly care, with an application to multiple other use cases. While single-modal systems provide a foundational understanding and an array of capabilities, the evolving landscape of multi-modal systems harnesses the synergies between different modalities, offering a more comprehensive perspective of human activities. Recent research~\cite{Fan2020}~\cite{memmesheimer2020} have showed that fusing data from multiple modalities can significantly improve the accuracy and robustness of activity detection.
From the recent literature, it can be seen that gait analysis has proven crucial in distinguishing health states using multiple sensors and algorithms.
The highlighted studies~\cite{Zhao2021}~\cite{Shao2022}, although focusing on specific conditions, reveal the potential for broader applications in elderly monitoring.
However, challenges persist, particularly in seamlessly integrating and processing diverse data streams without incurring exorbitant computational costs.
Future research should aim to improve sensor integration through the harmonization of data and the use of unified platforms, and the generalization across diverse populations, by designing more inclusive data collection plans~\cite{Bragazzi2022} and delving into stratified models.
Moreover, while a majority of research has been directed at detecting standard daily activities, there's an emerging emphasis on recognizing health-related movements and incidents specific to the elderly, such as falls~\cite{Li2018}. This not only enhances the safety for seniors living independently but also provides valuable insights to healthcare professionals for optimizing care regimens. In this context it would be interesting to see the evolution of HAR systems specifically tailored for specific demographics, such as the elderly. Future works may also consider the development of adaptive algorithms, capable of personalizing activity recognition based on individual behavior patterns (e.g., extracting from the activity diary~\cite{Fan2020}), thus paving the way for more personalized and effective elderly care solutions.

\subsection{Vital sign measurement}
\label{sec:vitalsignmeasurement}

\begin{table*}
\caption{Vital signs monitoring}
\label{tab:vitalSigns}
\begin{tabular}{|p{2cm}|p{4cm}|p{4cm}|p{6cm}|}
\hline
\textbf{Articles} & \textbf{Modalities}  & \textbf{Scenarios} & \textbf{Methods} \\ \hline

He~\textit{et al.}~\cite{He2022} & Xethru IR-UWB and Kinect, Vernier belt & respiration rate of 1-2 subjects \underline{sitting stationary}; classifying eupnea, Cheyne-Stokes respiration, Kussmaul respiration, and apnea & using the depth camera to detect and localize the subject; then, selecting the radar at the optimal location; random forest classifier  \\ \hline

Ren~\textit{et al.}~\cite{ren2017comparison} & Radar and optical camera based techniques & detecting vital signs such as respiratory rate (RR), heart rate (HR), and blood oxygen saturation in the \underline{sitting position}, \underline{varying illumination} & using \ac{UWB} stepped-frequency continuous-wave radar and imaging photoplethysmography (iPPG) techniques to measure vital signs \\ \hline

Yang~\textit{et al.}~\cite{yang2021remote} & FMCW radar and a digital camera &  Heart rate (HR) and breathing rate (BR), \underline{pre-exercise } and \underline{post-exercise} conditions, three subjects & the vital signs were extracted from the STFT spectrograms with the graph-based segmentation algorithm
\\ \hline


Xie~\textit{et al.}~\cite{Xie2021} & a pair of co-located UWB and depth sensors, medical devices (Nonin
LifeSense II and Masimo Pulse Oximeter) for ground truth & Heart rate and respiratory rate, \underline{sitting}, 1 - 3 subjects &  identifying four typical temporal and spectral patterns and a suitable RR/HR estimator for each, possible to monitor 3 subjects separating at only 20 cm \\ \hline

Shokouhmand \textit{et al.}~\cite{shokouhmand2022camera} & red-green-blue-depth (RGB-D) camera and radar &  respiratory rate (RR) and heart rate (HR), \underline{sitting} &  The camera estimates the human torso landmarks and a processing unit constantly adapts the radar beams to the direction of the subjects
\\ \hline

Yang~\textit{et al.}~\cite{Yang2020} & impulse-radio \ac{UWB} (IR-UWB) radar, optical and depth-sensing camera &  respiration rate, \underline{sitting}, 1 - 2 subjects &  face detection algorithm
based on the RGB video images, using the camera and utilizes the distance information acquired to find the subejcts and search the respiration waveform
\\ \hline

Chian~\textit{et al.}~\cite{chian2022vitalsigns} & Two radars and one thermal camera,  & respiration rates and heart rate, 1 - 5 subjects \underline{sitting}, Vernier respiration belt and Polar H10 as reference senros & Using a thermal camera to detect the number of people and their locations, identifying the respiration rates and heartbeat rates of multiple people.\\ \hline

Vilesov~\textit{et al.}~\cite{Vilesov2022} & TI AWR1443 FMCW radar and Zed2 RGB camera;  Philips MX800 clinical patient monitor for ground-truthing & heart rate prediction of 91 \underline{sitting} volunteers (28 light, 49 medium, and 14 dark skin), fairness in terms of similarity of outcome between the light and dark skin tone groups & Two-agent minimax problem: a fusion network mapping from the unimodal waveforms to the fused multimodal waveform and a discriminator minimizing the mutual information between the estimated waveform and skin-tone attribute \\ \hline




\end{tabular}
\end{table*}

Obtaining physiological data such as respiration rate and heart rate plays a crucial role in monitoring health and well-being of individuals.
In this section, we explore the non-invasive vital sign measurement techniques that utilize videos and radio frequency signals.

Unimodal vital-sign monitoring using vision or radio waves has been extensively studied in the literature~\cite{soto2022_wifivitalsign_survey, selvaraju2022_cameravitalsign_survey, zhang2022rf}. Soto~\textit{el al.}~\cite{soto2022_wifivitalsign_survey} provided a comprehensive survey of \ac{WiFi}-based vital-sign monitoring using \ac{CSI} information. Selvaraju~\textit{et al.}~\cite{selvaraju2022_cameravitalsign_survey} conducted a complete review of the recent advances in camera-based techniques for monitoring of five vital signs including heart rate, respiratory rate, blood pressure, body skin temperature, and oxygen saturation. Zhang~\textit{et al.}~\cite{zhang2022rf} discussed the overview of three types of electromagnetic radar systems such as continuous-wave, \ac{FM}, and \ac{UWB} to measure human physiological signs. These studies demonstrate the potential of one modality such as camera, WiFi signal, or radar sensor for non-contact vital sign monitoring in a variety of applications.

However, there are still several challenges to overcome under unimodal vital sign monitoring approaches. For example, systems based on radio-frequency signals suffer from signal interference and body motion artifacts.
On the other hand, camera-based systems are less accurate in poor-lighting conditions. Multimodal approaches can address such limitations and increase the reliability of non-contact vital sign monitoring~\cite{zhang2022rf}.
He~\textit{et al.}~\cite{He2022} combined depth cameras and radars to localize and estimate vital signs for one or two subjects.
The use of a 3D depth camera (Kinect) helps to ensure that the subjects being monitored are within the coverage area of three ultrawideband radars. These radars, which use impulse radio technology, are specifically designed for measuring respiration rate and detecting respiration patterns.
Yang~\textit{et al.}~\cite{Yang2020} also reported a system for monitoring simultaneously the respiration of two subjects with a hybrid camera-radar prototype. The system was built with an impulse-radio  ultra wide-band (IR-UWB) radar module and an optical and depth-sensing camera module. Subject detection is done by the camera and uses the distance information to guide the signal processing of the radar. Providing that context to the radar facilitates the extraction of respiration information. Instead of depth camera, Chian~\textit{et al.}~\cite{chian2022vitalsigns} proposed a system to track and identify the respiration rates and heartbeat rates of multiple people using a thermal camera and a Doppler radar. The thermal camera detects the number of people and their movements, and the doppler radars estimates the respiration and heartbeat. Other studies focused on the performance of the radar and optical techniques. For example, Ren~\textit{et al.} measured respiratory rate, heart rate and blood oxygen saturation with a combination of ultrawided stepped stepped-frequency continuous-wave radar and a camera for photoplethysmography (iPPG). In the study, they pointed out the differences of the modalities: The radar can be used for seeing-through walls, and optical technique is uniquely capable of measuring \(SpO_2\).
Yang~\textit{et al.}~\cite{yang2021remote} used a FMCW radar and a digital camera to measure the vital signs in real-life situations. The graph-based method provides HR measurements that are highly correlated with the golden standard device. Similarly, heart and respiratory rates can be obtained using a pair of co-located UWB and depth sensors identifying the spectral patterns \cite{Xie2021}.The work focuses in the robustness against harmonics and intermodulation. Likelihood of different respiration rate and heart rate are studied to detect corrupted signals. Shokouhmand~\textit{et al.}~\cite{shokouhmand2022camera} developed a system where the respiration rate and heart rate are monitored in real time by radar and a depth camera. While the depth camera tracks the locations of subjects, the radar estimates their respiratory and heart rates. A novel method is designed to optimize the regions of interest for monitoring the respiration rate and heart rate. A combination of micro Doppler motion effect and  photoplethysmography was used by Rong~\textit{et al.}~\cite{rong2022new} for cardiac pulse detection. The breath rate and heart rate are detected micro Doppler motion effect using  microwave frequency while the volumetric change of the heart is measured by photoplethysmography. In the study, they introduced the concept of terahertz-wave-plethysmography which detected blood volume changes in the upper dermis tissue.
Also implementing the fusion method for rPPG with a radar and a camera, Vilesov~\textit{et al.}~\cite{Vilesov2022} ensured the fairness in terms of similarity of the estimated heart rates between the light and dark skin tone groups. They formulated the problem as a two-agent minimax optimization procedure: a fusion network mapping from the unimodal waveforms to the fused multimodal waveform and a discriminator minimizing the mutual information between the estimated waveform and skin-tone attribute.

Two common challenging situations in non-contact vital sign measurement are body movement and multi-target monitoring.
The former can be tackled with movement reduction or cancellation by radars~\cite{cardillo2022vitalsign}~\cite{dai2022enhancement}~\cite{peng2021noncontact}.
The latter can be mitigated either by radars~\cite{wang2021multi}~\cite{feng2021multitarget} or by integrating cameras to track the targets and direct the radars towards the optimal region for measurement~\cite{shokouhmand2022camera}.

Recent research on non-contact vital sign measurements underscores the evolution from unimodal to multimodal monitoring approaches. While camera-based techniques offer depth perception and subject localization, their performance often diminishes under poor lighting conditions. On the other hand, radio-frequency methods, though robust in various environmental conditions, grapple with interference from body motions and other electronic devices. We can particularly highlight the emerging trend of integrating visual and RF modalities~\cite{shokouhmand2022camera}~\cite{chian2022vitalsigns}~\cite{He2022}~\cite{Vilesov2022}, which serves to counteract the individual limitations of each method. Such fusion notably enhances the system's ability to navigate challenges, especially in situations involving body movement and multi-target monitoring. In this context, future efforts might benefit from developing advanced algorithms for seamless integration of multiple data sources, investigating the impact of environmental factors on multimodal systems and exploring the potential of newer sensing technologies.

\subsection{User identification}
\label{sec:identification}

\begin{table*}
\caption{User identification}
\begin{tabular}{|p{3cm}|p{3cm}|p{4cm}|p{6cm}|}
\hline
\textbf{Articles}  & \textbf{Modalities}  & \textbf{Features}    & \textbf{Methods \& Results} \\ \hline
Shi~\textit{et al.}~\cite{Shi2023} & mmWave radar, RGB camera & Fusing features from gait energy images and time-Doppler spectrograms  &  121 subjects (72 men and 49 women) with eight views and three walking conditions, experimenting with different fusion methods (voting-based decision fusion, feature concatenation, and feature fusion based on learned weights) to show that concatenating features along the along the space and temporal dimensions achieved the best accurary (up to 95.4\%) \\ \hline
Li~\textit{et al.}~\cite{li2016id} & RFID, Kinect & RFID phase features (velocity, average phase change,\dots), RSSI features, Kinect skeleton motion features  & ranking correlation between bodies Kinect and tags of the RFID reader, SVM classifier, assigning IDs to individuals within 4 seconds with 96.6\% accuracy \\ \hline
Chen~\textit{et al.}~\cite{Chen2022_RFCam} & WiFi, RGB camera & WiFi CSI and videos, device AoA using phase differences between antennas, LSTM & vision-assisted calibration and Bayesian inference to improve the
device AoA estimation \\  \hline
Cao~\textit{et al.}~\cite{Cao2022} & WiFi, RGB camera & PointNet-based learned features from CSI and RGB & PointNet with attention mechanism, LSTM encoder-decoder network \\  \hline
Liu~\textit{et al.}~\cite{Liu2022}& WiFi, RGB camera & WiFi Fine Timing Measurements, inertial data, and RGB-D videos & LSTM as feature extractor to generate motion trajectories, LSTM as feature extractor, Maximum Matching Weighted
Bipartite Graphs for matching track and phone IDs \\  \hline
Deng~\textit{et al.}~\cite{Deng2022_Gait} & WiFi, RGB camera & CSI, Lightweight Residual Convolution Network for feature learning & LRCN, LSTM, feature fusion, 94.2\% recognition accuracy \\  \hline
Fang~\textit{et al.}~\cite{Fang2020} & WiFi, panaromic camera & WiFi CSI, panaromic camera & estimating the Angle of Arrival (AoA) of the smartphone, cross-modal trajectory matching to determine the identity of the individuals \\ \hline
Luchetti~\textit{et al.}~\cite{Luchetti2021} & UWB, RGB camera & Locations & UWB device and camera to track tags carried by users \\ \hline
\end{tabular}
\end{table*}

This section focuses on techniques and systems that leverage multimodal data to infer the user identities across all modalities.
Users can be identified using gait analysis, which employ various sensors such as cameras, pressure sensors, radars, and accelerometers~\cite{Wan2018}~\cite{Nambiar2019}.
An identification model maps a feature vector $v \in \mathbb{R}^k$ to an identity.
Machine learning methods can be leveraged to aligning and transforming features of one modality to another~\cite{Moghaddam2023}.

Due to their ubiquity in indoor environment, WiFi signals are combined with visual information for user identification.
Deng~\textit{et al.}~\cite{Deng2022_Gait} utilized WiFi signals and videos for human identification.
They proposed a two-stream network architecture based on the Lightweight Residual Convolution Network (LRCN) to learn features for WiFi and video data. The learned features were concatenated as inputs to a LSTM model to recognize users with an accuracy of 94.2\%.
The EyeFi system of Fang~\textit{et al.}~\cite{Fang2020} integrated a WiFi chipset into a panaromic overhead camera in order to combine motion trajectories from both vision and RF modalities for human identification.
They trained a neural network based on a student-teacher model to estimate the Angle of Arrival (AoA) of WiFi packets from the CSI values.

In some scenarios, one user moves from a camera-allowed region to a camera-restricted area.
This problem requires solutions to find the alignment between two or more modalities.
To address the cross-modal human re-identification problem, Cao~\textit{et al.}~\cite{Cao2022} extracted point cloud data from both modalities and match them.
They utilized a PointNet-based model to extract features from CSI and RGB.
Their PointNet architecture integrated the attention mechanism and a LSTM encoder-decoder network to implement user re-identification across two modalities.
Li~\textit{et al.}~\cite{li2016id} extracted RFID phase features (such as velocity and average phase change), RSSI features, and Kinect-based skeleton features for data association.
Their method ranked the correlation between the body motion generated by Kinect and RFID tags
The system could assign IDs to individuals with the accuracy of 96.6\% accuracy.
Liu~\textit{et al.}~\cite{Liu2022} leveraged WiFi Fine Timing Measurements (FTM) and inertial measurement unit (IMU) sensor data captured by the user's smartphone to associate the user detected on a camera footage with their corresponding smartphone identifier (e.g. WiFi MAC address), using a recurrent multi-modal deep neural network.
Chen~\textit{et al.}~\cite{Chen2022_RFCam} used a multi-antenna WiFi radio and a camera to monitor users in the area.
The approach processed WiFi CSI data of WiFi packets to extract features on locations, motion, and activities of users.
It fused these features with visual information to find the best matches of device carriers in videos with network IDs of the phones.
Liu~\textit{et al.}~\cite{Liu2022} analysed WiFi Fine Timing Measurements, inertial data, and RGB-D videos to generate motion trajectories, using a LSTM model.
Then, they implemented Maximum Matching Weighted Bipartite Graphs in order to match visual tracks and phone IDs.
Luchetti~\textit{et al.}~\cite{Luchetti2021} leveraged UWB devices and cameras to track tags carried by users.

The integration of multimodal data for user identification showcases the potential of synergizing diverse modalities, with WiFi signals and visual cues being at the forefront. Current methodologies predominantly employ machine learning techniques, from LSTMs to PointNet architectures, to align~\cite{li2016id}~\cite{Liu2022}, transform~\cite{Chen2022_RFCam}, and extrapolate features~\cite{Fang2020} from one modality to another, achieving impressive identification accuracies. However, challenges such as transitions from camera-enabled to camera-restricted zones, interference, and real-time processing are still largely unaddressed. Future research should focus int the seamless fusion of data, addressing privacy concerns, and adapting these techniques to heterogeneous environments and ensuring consistent performance across varying conditions.

\subsection{User localization}
\label{sec:localization}

\begin{table*}
\caption{User localization}
\label{tab:Localization}
\begin{tabular}{|p{3cm}|p{2cm}|p{3cm}|p{8cm}|}
\hline
\textbf{Articles}  & \textbf{Modalities} & \textbf{Features}  & \textbf{Methods \& Results} \\ \hline
Suwannaphong~\textit{et al.}~\cite{suwannaphong2022radio} & Bluetooth, Camera & BLE RSSI (camera for ground-truth) & RSSI augmentation is done from multiple BLE devices using camera ground-truths and one-shot learning \\ \hline
Jiang~\textit{et al.}~\cite{Jiang2022} & Bluetooth, Camera & BLE 5.1, visual point cloud, and IMU & locating the BLE transmitter in the camera coordinate, homography-based matching mechanism to obtain the depth information of the target BLE device, 3.4$\deg$ angular error and 8.4cm position error in median, outperforming the state-of-the-art by 48\% and 8\%, respectively; overall accuracy over 90\% \\  \hline
Ishihara~\textit{et al.}~\cite{Ishihara2018} & Bluetooth, Camera & CNN ouput & image-based localization by integrating Bluetooth radio-wave signal readings, using a dual-stream CNN \\ \hline

Deng~\textit{et al.}~\cite{Deng2022} & Radar, Camera & mmWave Radar and Camera Fusion & Object detection on
drones \\ \hline
Li~\textit{et al.}~\cite{Li2022_Object} & LiDAR, Radar, Camera & LiDAR, Radar, and camera for object detection and tracking & multiple modality combinations, LiDAR: point-wise concantenation, Radar: pillar feature encoder, camera: image feature encoder \\ \hline
Li~\textit{et al.}~\cite{Li2022_Pedestrian} & Radar, Camera & mmWave radar IWR6843ISK-ODS and RGB camera & distingushing living pedestrian and portrait billboard, a feature fusion network of mmWave radar and computer vision based on attention mechanism, mAP of 97.7\% \\ \hline

Papaioannou~\textit{et al.}~\cite{Papaioannou2014} & WiFi, Camera & WiFi-based and Camera-based trajectories & Using WiFi measurements to merge camera-based tracklets based on tracklet trees \\  \hline
Zhao~\textit{et al.}~\cite{Zhao2020} & WiFi, Camera & WiFi-based and Camera-based trajectories & Using WiFi CSI to aid distance calculation based on inertial data, then combining it with vision-based localization, YOLO V2 for target detection \\  \hline
Cai~\textit{et al.}~\cite{Cai2020} & WiFi, Camera & Trajectories & Vision-based tracking to extract the object trajectory,  radio signals identify anonymous visual tracks and correct errors \\  \hline
Xu~\textit{et al.}~\cite{Xu2019_IVR} & WiFi, Camera & Visual data, radio (RSS), and inertial data to extract trajectories & Augmented particle filters, Trajectory fusion, localization accuracy of 0.7m, outperforming the state-of-the-art systems by >70\% \\ \hline

Zhang~\textit{et al.}~\cite{Zhang2021} & UWB, Infrared Camera &  Infrared on-board camera locating by identifying artificial landmarks attached to the ceiling, TOA-based UWB positioning method & Asymmetric Double-sided Two-way ranging, Extended Kalman filter (EKF) is used to fuse the real-time location information \\ \hline
Nguyen~\textit{et al.}~\cite{Nguyen2021} & UWB, Infrared Camera, IMU & Trajectories from camera, IMU, and UWB & Monocular camera, a 6-DoF IMU, and a single \ac{UWB} anchor for robot localization \\ \hline
Liu~\textit{et al.}~\cite{Liu2020} & UWB, Camera & Trajectories from UWB and monocular camera & EKF (extended Kalman filter), resolving scale ambiguity and repositioning of the monocular ORB-SLAM, indoor positioning accuracy of 0.2m \\ \hline

Varotto~\textit{et al.}~\cite{Varotto2021} & Radio receiver, Camera & Target locations & Drone with a radio receiver and a camera for detecting a radio-emitting target, recursive Bayesian Estimation scheme that uses camera observations to refine radio measurements \\ \hline

Streubel~\textit{et al.}~\cite{streubel2016fusion} & Indoor Pedestrian Tracking & Camera and Radar (MIMO-FMCW) & Fusion of stereo camera and radar targets to track pedestrians  \\ \hline
Pearce~\textit{et al.}~\cite{pearce2022combined} & Tracking human movement patterns & mmWave and Camera (as supervision signals) & Framework for training a mmWave radar model with a camera for labeling the data and supervising the radar model   \\
\hline
Lim~\textit{et al.}~\cite{lim2021radical} & Human detection and tracking & FMCW Radar, Depth, IMU, and RGB Camera Data &  3D convolution based supervised regression model, Out-of-range pixel classification model\\
\hline
Cai~\textit{et al.}~\cite{Cai2023} & Human detection and tracking & mmWave radar and thermal camera & Uncertainty-guided fusion framework using Bayesian Neural Network to jointly extract features for for human detection in visual degradation conditions \\ \hline
\end{tabular}
\end{table*}

Within the context of indoor human centered monitoring systems, one of the essential tasks is indoor localization.
An localization system aims to estimate the user location $x \in \mathbb{R}^3$ using a feature vector $v \in \mathbb{R}^k$.
User positions and movement trajectories $t = [x_1,\dots,x_i,\dots, x_n ]$ provide useful measurements on the well-being of the individuals living at their home.

Multimodal data improves localization in challenging situations such as cluttered or dark environments in indoor scenarios.
For example, in a factory, surveillance cameras can be obstructed by equipment.
Tarkowski~\textit{et al.}~\cite{tarkowski2016wireless} proposed a hybrid localization system combining vision-based, radio-based and inertial techniques to solve localization issues in difficult and complicated industrial situations.
Similarly, Woznica~\textit{et al.}~\cite{woznica2014rf} and Wang~\textit{et al.}\cite{wang2011rfid} tackled the problem of localization using multimodal analysis of radio and vision.
In a dim lighting museum setting with multiple people moving around exhibit stands, Papaioannou~\textit{et al.}~\cite{Papaioannou2014} integrated WiFi measurements to concatenate visual tracklets into accurate and continuous paths to reduce the localization error to below 1 meter.
Zhao~\textit{et al.}~\cite{Zhao2020} enhances indoor localization of users with multimodal sensing: camera, wearable IMU sensors, and ambient CSI data of WiFi signal.
They estimated distance and orientation in indoor environments to achieve sub-meter accuracy.
Zhang~\textit{et al.}~\cite{Zhang2021} introduced a positioning method that combined infrared vision and ultra-wideband radio device.
The former identified landmarks attached to the ceiling while the latter used an adaptive weight positioning algorithm to improve the accuracy at the edge of the UWB coverage area.
Extended Kalman filter (EKF) was used to fuse their location information.
Xu~\textit{et al.}~\cite{Xu2019_IVR} introduced an multi-sensor localization system with an accuracy of 0.7m in indoor environments.
The sensors included: cameras, RSS of wireless signals, and wearable inertial sensors.
iVR used an augmented particle filter to combine multiple unimodal locations.
Jiang~\textit{et al.}~\cite{Jiang2022} deployed BLE 5.1, visual point cloud, and IMU to reduce the positioning error to 3.4$\deg$ angular error and 8.4cm position error in median, which outperformed the state-of-the-art by 48\% and 8\%, respectively.
Liu~\textit{et al.}~\cite{Liu2020} utilized UWB and monocular camera to resolve scale ambiguity and repositioning of the monocular ORB-SLAM.
In an retail environment, Sturari~\textit{et al.}~\cite{Sturari2016} proposed a sensor fusion system consisting active radio beacons and RGB-D cameras. This system was used for customer indoor localization and tracking.
A mobile application received (with a frequency of 5Hz) raw data (beacon MAC address, broadcasting power, and RSSI) to get the device position.
On the other hand, the RGB-D cameras monitored an area to track people and propagate position data.
Following the research direction of combining visual and radio data, other researchers leverage specially-designed equipment for sensing: radar.
Streubel~\textit{et al.}~\cite{streubel2016fusion} combined camera and radar features to detect and track indoor pedestrians. They have taken stereo camera features and multiple input multiple output (MIMO) frequency modulated continuous wave (FMCW) radar features, and then used a standard linear Kalman filter based method for tracking the humans, and have shown how using a heterogenous sensor fusion gives better performance when compared with the sensors individually.
The visual richness of videos can be exploited to enhance radar systems.
Pearce~\textit{et al.}~\cite{pearce2022combined} leveraged camera data to train an mmWave radar system to track and classify human movement patterns.
Although RGB cameras have achieved great performance in human detection and tracking, they are susceptible to visual degradation.
To address this issue, Cai~\textit{et al.}~\cite{Cai2023} introduced a uncertainty-guided fusion framework based on a Bayesian Neural Network to combine thermal videos and mmWave radar signals for detecting multiple users.

Different modalities can be combined in a parallel manner to improve the performance of object detection and localization systems.
Ishihara~\textit{et al.}~\cite{Ishihara2018} enhanced image-based localization accuracy by integrating Bluetooth radio-wave signal readings, using a dual-stream CNN.
Qiu~\textit{et al.}~\cite{Qiu2022detection} used the radio localization and identifier information from the radio signals to assist the human detection.
On the other hand, one modality can be leveraged to increase the measurement accuracy of the other.
For instance, Varotto~\textit{et al.}~\cite{Varotto2021} detected a radio-emitting target with an aerial robot equipped with a radio receiver and a camera. They implemented a Recursive Bayesian Estimation scheme that utilized camera observations to refine radio measurements.
Deng~\textit{et al.}~\cite{Deng2022} introduced Geryon using camera and mmWave radar on drones for object detection.
The visual information assisted radar to reduce the aggravated sparsity and noise of point-cloud data.
On the other hand, a radar-based saliency extraction algorithm decreased bandwidth consumption and offloading latency of camera.
Beyond camera and radar combination, Lim~\textit{et al.}~\cite{lim2021radical} provided an FMCW radar dataset that is synchronized with RGB-D and also inertial measurement unit (IMU) measurements. It includes indoor scenes of different rooms and including different numbers of people and stationary obstructions, and the authors have built models to perform human tracking with radar and depth information. Compared to traditional RGB-D cameras, they have shown that using radar results in far lesser noise and better understanding of varying velocity.

Vision-based tracking offers high localization accuracy but suffers from identity switching and fragmentation errors when multiple users are in close to each other or occluded.
On the other hand, radio-based tracking is reliable in object detection and identification but less accurate in localization.
Cai~\textit{et al.}~\cite{Cai2020} leveraged radio-based identification and localization techniques to augment anonymous visual tracks with identity information and enhance the accuracy of vision-based object detection.
Nguyen~\textit{et al.}~\cite{Nguyen2021} deployed a monocular camera, a 6-DoF IMU, and a single unknown Ultra-wideband (UWB) anchor to implement accurate and drift-reduced localization.
They processed the UWB measurements by leveraging the propagated data computed by the visual-inertial-odometry (VIO) pipeline.

The recent literature underscores the significance of multimodal data in enhancing indoor localization, particularly in complex scenarios like cluttered spaces or areas with low light. The synergy of different modalities, be it Bluetooth and cameras~\cite{suwannaphong2022radio}~\cite{Jiang2022}, radars and LiDARs~\cite{Li2022_Pedestrian}, thermal camera and mmWave radar~\cite{Cai2023}, or UWB, camera, and inertial sensors~\cite{Nguyen2021}, offers a robust solution to limitations inherent to each individual system. Despite the recent effort, some challenges still persist. Vision-based systems present obvious problems in occluded spaces while radio-based system still show a relatively low accuracy and the combination approaches are tailored according to modalities and use cases.
An emerging and promising research direction, self-supervised learning, could be exploited by having certain modalities aid in the annotation of others, such as computer vision models providing labels for radio-based systems. This could result in larger-scale databases with more variability on scenarios and modalities, potentially improving the reliability in dynamically-changing environments.

\subsection{Depth estimation}
\label{sec:3d}

\begin{table*}
\caption{Depth estimation}
\begin{tabular}{|p{2cm}|p{5cm}|p{4cm}|p{5cm}|}
\hline
\textbf{Articles}                           & \textbf{Task}    & \textbf{Modalities} & \textbf{Methodology} \\ \hline

Xu~\textit{et al.}~\cite{xu2022learned} & Depth Estimation for indoor Mapping & mmWave and Lidar (as supervision signals) &  3D convolution based supervised regression model, Out-of-range pixel classification
mode\\ \hline
Lu~\textit{et al.}~\cite{lu2020see} & Scene understanding in smoke-filled environments & Cross-modal supervision of LiDAR to radar data & Conditional-GAN-based approach, classifying obstacles on the reconstructed grid map (such as door, lift, wall, and glass)
\\ \hline
Long~\textit{et al.}~\cite{long2018fusion} & Obstacle detection in diverse illumination conditions & RGB-Depth data with mmWave radar data & Particle filter data fusion to reduce uncertainties and obtain more accurate state estimation \\ \hline
Ding~\textit{et al.}~\cite{Ding2023} &  3D human mesh estimation & RGB/IR images and mmWave point clouds & Combining outputs of PointNet, ShapeNet, and CamNet as inputs to a vertex-based Skinned Multi-Person Linear Model to estimate body shapes  \\ \hline


\end{tabular}
\end{table*}

In this section, we discuss the methods and applications which use multimodal data for depth estimation and related tasks such as 3D modelling and scene understanding. Depth estimation is useful to enhance the environmental perception capability of the intelligent machines in monitoring and surveillance tasks. Several works have utilized multimodal data for understanding scenes and objects in the 3D setting~\cite{xu2022learned}\cite{lu2020see}\cite{long2018fusion}.
Combining the discussed multimodal approaches are useful estimating in depth related tasks. Modalities such as radio along with RGB data help in aiding the perception of depth which is a crucial step for the efficient understanding of human-centered 3D indoor environments.

The abilities of radio signals to provide information invariant to color and lighting condition make it a useful modality, when being combined with RGB features such as color, shape, and textual information.
For instance, Long~\textit{et al.}~\cite{long2018fusion} proposed a data fusion method based on Particle Filter, which effectively utilized RGB-Depth data with mmWave radar data to accurately detect multiple objects with varying distances and orientations, even in diverse illumination conditions.

Cross-modal supervision is another approach to leverage the strengths of different modalities.
Xu~\textit{et al.}~\cite{xu2022learned} present a depth map regression model for imaging radar perception by using mmWave radar with LiDAR based supervision. Their approach produces depth maps that visually match LiDAR ground truth, and it is shown how their model successfully detects elements like floors and ceilings that are mostly undetectable and are crucial for indoor mapping tasks. Some challenge discussed in this work includes a lack of understanding for glass objects, which the authors believe could be mitigated by performing better understanding of materials, and also some noisy near-range observations and due to the complicated reflection path. Apart from complications due to obstructions and material hindrances, there can also be situations when smoke and dust obscure vision, rendering camera and laser based devices inoperable. In a similar combination, to understanding the objects and obstacles in indoor mapping, Lu~\textit{et al.}~\cite{lu2020see} proposed a solution, called milliMap, to tackle the scene understanding challenges in smoke-filled environments. Their conditional-GAN-based approach handled sparsity and multi-path noise of mmWave radar signals by combining cross-modal supervision from a LiDAR and the strong geometric priors of indoor spaces.
Along with indoor mapping, their approach also classified the semantics of the obstacles on their reconstructed grid map (such as door, lift, wall, and glass) which improved the scene reconstruction accuracy.

Reconstructing the 3D model of a human body is essential in human monitoring systems.
In this application, multimodal data has showed their advantages since multiple modalities such as RGB/IR cameras and mmWave radars can complement each other.
For example, Ding~\textit{et al.}~\cite{Ding2023} fused features from RGB/IR images and mmWave radar point cloud to estimate a 3D human body mesh.

The potential of multimodal data sources, focusing on RGB, radar, and other radio devices, is still largely unexplored in the field of depth estimation~\cite{xu2022learned}, 3D modeling~\cite{Ding2023}, and scene understanding~\cite{lu2020see}~\cite{long2018fusion}. The combination of RGB color and the depth awareness of radar and radio signals could potentially offer a solution to conventional challenges such as obstructions, occlusions, or variable illumination~\cite{long2018fusion}. However, while significant progress has been made in indoor mapping, some challenges persist. These include occlusions by walls, difficulties in understanding glass objects, noise in near-range observations, and complications from partially transparent substances such as smoke and dust~\cite{xu2022learned}\cite{lu2020see}. There is an immense potential in analysis of multi-path propagation effects of radio-waves, which could be used by enhanced fusion techniques that complement cameras and dealing with ambiguous monocular information. 

\section{Multimodal methods}
\label{sec:methods}

\begin{figure}
     \centering
     \begin{subfigure}[b]{0.4\textwidth}
         \centering
         \includegraphics[width=\textwidth]{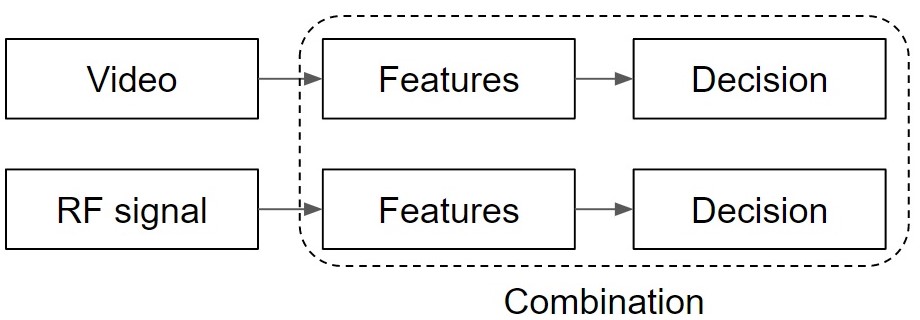}
         \caption{Combination}
         \label{fig:combination}
     \end{subfigure}
     \begin{subfigure}[b]{0.4\textwidth}
         \centering
         \includegraphics[width=\textwidth]{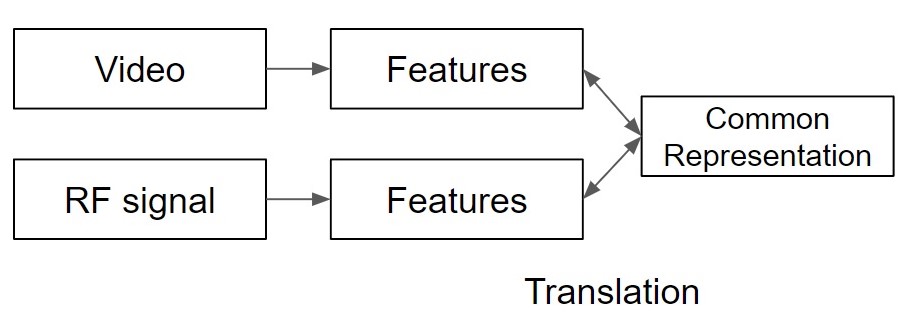}
         \caption{Transformation}
         \label{fig:translation}
     \end{subfigure}
     \begin{subfigure}[b]{0.4\textwidth}
         \centering
         \includegraphics[width=\textwidth]{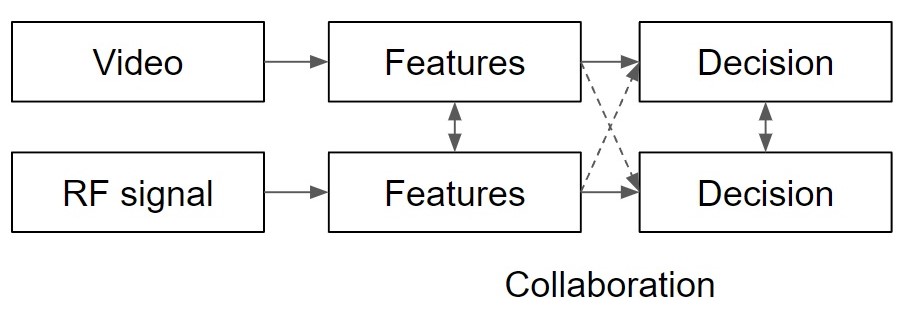}
         \caption{Collaboration}
         \label{fig:collaboration}
     \end{subfigure}
    \caption{Categorization of multimodal approaches}
    \label{fig:multimodal_approaches}
\end{figure}

Within the ambit of multimodal monitoring systems, we present a distinct categorization of approaches, as illustrated in Figure~\ref{fig:multimodal_approaches}. The categories include:
\begin{enumerate*}
\item the combination of various sensing modalities to yield a unified output,
\item the transformation of one modality into another, and
\item the mutual collaboration between different modalities, where each one aids in refining the other.
\end{enumerate*}
Historical reviews and surveys~\cite{Baltruvsaitis2018Survey}\cite{Guo2019} predominantly shine light on the first two approaches—combining and transforming multimodal data. The essence behind these techniques is to distill an optimal representation from the data conglomerate. Diverging slightly, we delve deeper into the notion of collaboration. In this paradigm, one modality's data can influence and optimize another's sensor operations, thereby enhancing the system's overall performance. This intriguing interplay among modalities, especially collaboration, will be further elaborated upon in the subsequent sections.

\subsection{Multimodal fusion}
\begin{figure*}
\includegraphics[width=\textwidth]{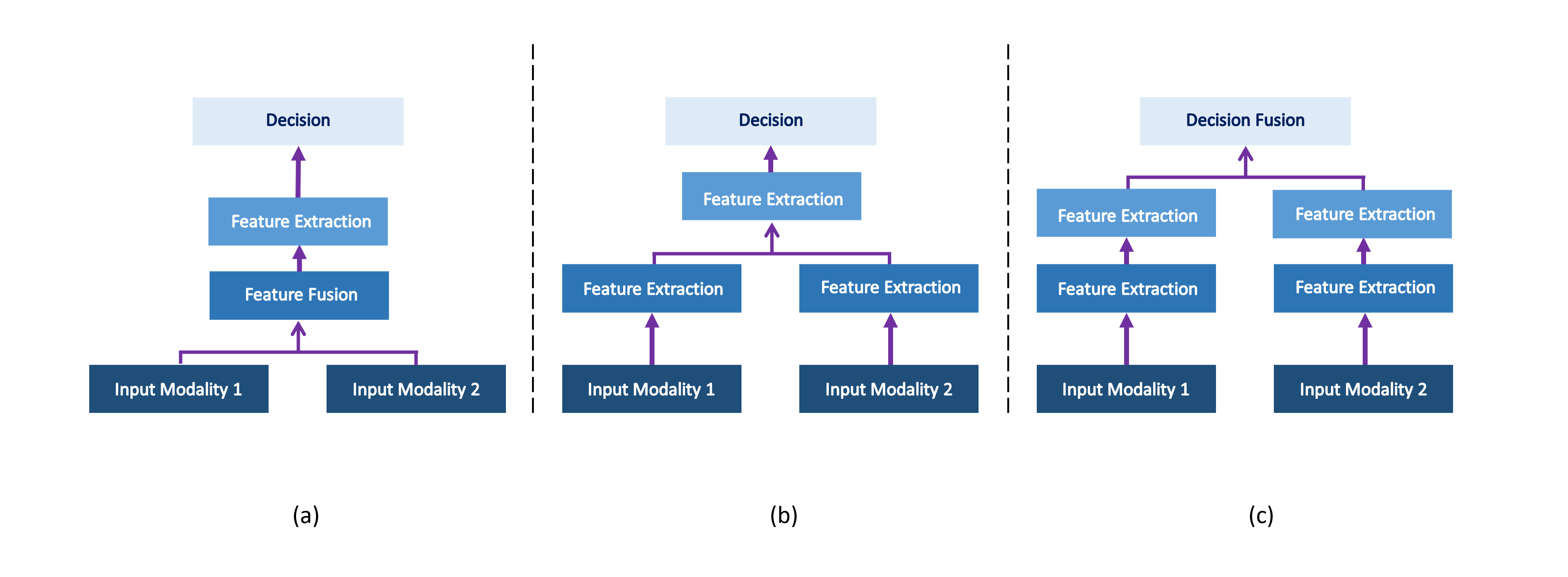}
\caption{Combination methods: (a) Early Fusion (b) Intermediate Fusion (c) Late Fusion}
\label{fig:featureFusions}
\end{figure*}

Multimodal fusion methods combine visual data and radio signals to improve the performance of machine learning models (see Figure~\ref{fig:featureFusions}) in terms of accuracy and reliability.
For example, Qiu~\textit{et al.}~\cite{Qiu2022detection} performed a radio position information-aided human detection to enhance the reliability and accuracy of human camera-based detection.  
Krahnstoever~\textit{et al.}~\cite{krahnstoever2005activity} augmented vision-based human motion tracking with radio frequency identification (RFID) technology. In a similar line of work, Sturari~\textit{et al.}~\cite{sturari2016robust} developed a sensor fusion system of active radio beacons and RGB-D cameras for investigating shopper movements and behavior in retail environments. Stotko~\textit{et al.}\cite{stotko2019albedo} used RGB-D and IR for reconstructing geometry and the spatially varying surface albedo of a scene. Similarly, Muaaz~\textit{et al.}~\cite{Muaaz2020} perform human activity recognition by combining Wi-Fi (CSI) with wearable inertial measurement unit (IMU) in Android phones. Qin~\textit{et al.}~\cite{qin2020imaging} fused time series data from different wearable sensors to recognize human activity with a residual network. In a attempt to generalize action recognition across various modalities, Memmesheimer~\textit{et al.}~\cite{memmesheimer2020} experimented with four datasets containing CSI informatoin from WiFi devices, skeleton information from camera and inertial, and body movements from a motion capture system.

The literature has started to explore the diverse fusion techniques that integrate vision and wireless communication technologies for a more accurate prediction of activities and scenes~\cite{Qiu2022detection}. By combining various sensing modalities, such as infrared images, radio position data, and RGB videos, we can achieve more robust results, especially in challenging settings such as low-light conditions. The intersection of tools, from RGB-D cameras to Wi-Fi CSI and wearable sensors, can enrich our understanding of multimodal fusion. However, there are still challenges in seamlessly merging data from different sources while maintaining consistent performance in different situations. Future research should focus on refining fusion methods, enhancing the richness of datasets, and creating systems that can adjust and learn in real-time to better capture and interpret dynamic indoor environments and human actions.

\subsection{Transformation}
When working with information from various sources, one task is to transform or map data collected with one sensing modality to that of others.
This is essential in a multi-sensor elderly monitoring system since we need to match feature vectors between different modalities.
For example, it is possible to install a camera in a living room but that would violate the user in a bedroom. Instead, a non-vision modality such as a radar should be deployed. A transformation technique is useful to map visual data to radar data in order to identify one user moving between these two rooms.

\subsubsection{Non-learning transformation}
Non-learning approaches rely on domain knowledge to map data collected by different sensor types (i.e. different modalities) to the same representation.
Different types of sensors can capture data in the same format.
For example, point cloud can be extracted from both radio frequency signals and videos.
Cao~\textit{et al.}~\cite{Cao2022} utilized PointNet-based models to extract point cloud data from WiFi CSI and RGB features.
They matched the data points in both modalities for user re-identification.
Multivariate signals can be plotted, saved as images, and processed by convolutional neural networks (CNNs).
Memmesheimer~\textit{et al.}~\cite{memmesheimer2020} transformed inertial data, and WiFi CSI into 2D images as input to a CNN-based classifier.
Another solution is to select one modality as the~\textit{anchor} space which other modalities project their data into.
Dimitrievski~\textit{et al.}~\cite{dimitrievski2019tracking} first used the Fast-CNN vision detector to generate a set of bounding boxes around detected objects. Then, they projected the boxes into the range-azimuth space of the radar data.
Similarly, Cui~\textit{et al.}~\cite{Cui2021tracking} generated bounding boxes with YOLOv3 on the image plane.
After that, they projected the detections from the pixel coordinate to the radar coordinate.
The non-learning techniques rely on domain knowledge and they need to be customized for different sets of sensors.

\subsubsection{Learning-based transformation}
Learning techniques rely on neural networks such as encoder-decoder to learn the common representation of different modalities.
Autoencoder is well-known for unsupervised representations learning.
It contains two component: one is an encoder and the other is a decoder.
The encoder transform the input feature vector into a compressed representation (i.e., latent representation).
The decoder reconstructs the input from this latent representation in such a way that the reconstruction error is minimized.
Multimodal user identification involves the transformation of data from multiple sources to a common form.
Cao~\textit{et al.}~\cite{Cao2022_ViTag} introduced ViTag in order to link a sequence of visual bounding boxes with inertial data and Wi-Fi Fine Time Measurements (FTM) from smartphones.
The system used a multimodal LSTM encoder-decoder network to perform cross-modal translation that reconstructed IMU and FTM readings from camera-based bounding boxes.
Then, the transformed output was matched with the observed data.
Piechocki~\textit{et al.}~\cite{Piechocki2023} learned a common latent representation of videos and WiFi signals for activity recognition.
Especially, their generative model can reconstruct data of a missing modality.
These transformation models can be considered as a universal solution for any sensors but they require a significant amount of data for training.

In the exploration of the literature on data transformation within multi-sensor systems, two distinct strategies emerged: non-learning and learning-based techniques. The non-learning techniques rely mostly on domain knowledge, converting sensor data into shared representations by leveraging inherent similarities between modalities or designating one modality as the anchor space. While effective, they demand customization for different sensor combinations, posing a challenge to scalability. Conversely, learning-based transformations propose the utilization of neural network architectures, such as autoencoders, to learn shared-modality representations. While they offer more universal applicability, their dependence on vast training datasets remains a limitation. Moving forward, the research community could focus on creating hybrid transformation models that merge domain knowledge with adaptive learning to ensure robust, scalable, and efficient data transformation in dynamic multi-sensor environments.

\subsection{Collaboration}
\label{sec:collaboration}

In recent years, the concept of modality collaboration has emerged as a possible approach to enhance the performance and reliability of multi-sensor systems. By leveraging the strengths of individual sensing modalities, researchers aim to address their inherent limitations, using collaborative schemes with other sensors. This collaboration manifests in two primary forms: co-learning and cross-modal guidance, both of which exploit synergies between modalities to deliver enhanced outcomes in a variety of applications.
Co-learning is about the transfer of learned knowledge between modalities.
In cross-modal guidance, one modality can control another in order to enhance the performance of the whole multimodal system.

\subsubsection{Co-learning}
Co-learning through multiple modalities can be useful to perform robust and reliable action recognition in non-contact human assisting systems. For example, using vision alongside radio waves as the prior knowledge can be helpful to overcome the shortcomings of radio signals. Similarly, radio signal data can be useful to assist vision-based sensors and perform reliable activity recognition under low illumination, dark, and occlusion conditions. 
Zhao~\textit{et al.}~\cite{zhao2018through} estimated human pose through walls and occlusions using cross-modal teacher-student network that transferred the visual knowledge of human pose from the video-based teacher network and supervised the RF-based student network to predict keypoint confidence maps. Bocus~\textit{et al.}~~\cite{bocus2022uwbdataset} proposed a ultra wideband dataset (UWB) for human activity recognition (HAR) built with COTS RF device input data and camera ground truth. They used commercial off-the-shelf UWB-equipment to generate the dataset against camera ground truths. Human pose-estimation using radio frequency signals with the help of camera ground-truth have been studied by several research groups~\cite{xie_humanpose_2022}\cite{sengupta_humnpose_2020}\cite{jiang_humanpose_2020}. Zou~\textit{et al.}~\cite{zou2019wifi} introduces WiVi which identifies common human activities in an accurate and device-free manner using commercial WiFi-enabled IoT devices and camera, and by performing multimodal fusion at the decision level to combine the strength of these modalities. Fei~\textit{et al.}~\cite{wang2019wifiperception}, proposed a WiFi-based human perception or detection with camera-based annotations.

Co-learning methodologies have paved the way for enhanced action recognition in non-contact human assisting systems. The most typical approaches have focused on integrating visual cues with radio wave insights. However, challenges such as efficient multi-modal data integration and the optimization of cross-modal translation techniques still remain, while adaptive fusion strategies could be a reasonable future research direction.

\subsubsection{Cross-modal supervision} The concept of cross-modal supervision stands out as a method where researchers leverage the unique strengths of one modality to enhance the performance of other modalities in a task.

Considering the challenge of labelling mmWave radar data, Pearce~\textit{et al.}~\cite{pearce2022combined} have proposed a framework to train a mmWave radar with a camera that solves the problem and helps in annotating the radar data. The work included recordings in indoor scenario, and uses camera and radar in a teacher-student fashion, which they show can be useful in tracking and classifying human movement patterns.
Shokouhmand~\textit{et al.}~\cite{shokouhmand2022camera} introduced a camera-guided radar to monitor vital signs. An RGB-D camera detected the human torso landmarks which was used to steer the radar beams to the direction of the subjects.
In multimodal localization, Qiu~\textit{et al.}~\cite{Qiu2022detection} utilized a radio-based positioning technique to improve the reliability and accuracy of human camera-based detection.
They leveraged the region proposals generated from radio localization to suppress the false detections and reduce miss detections.
On the other hand, He~\textit{et al.}~\cite{He2022}\cite{He2020} implemented an hybrid approach that used both modalities for localization and leveraged visual data to select the optimal radar for vital sign measurement.
Another approach is to use a well-established model to train another model (on another modality).
Zhao~\textit{et al.}~\cite{zhao2018through} designed a cross-modal teacher-student network which transferred the visual knowledge of human pose to radio-frequency signals.
Two or more modalities can supplement each other in various scenarios. 
Human assisting systems that integrate cameras and radio devices have showed their effectiveness in various application domains.
Radio waves aid visual data in scenarios that hinder the operations of cameras, e.g. occluded or private areas \cite{zhao2018through}.
One of the important challenges in tracking humans in indoor settings is occlusion of the subjects by walls. Using only camera device identifying the movement patterns and pose becomes difficult in such scenarios. The ability of radar to pass through walls help in tackling the issue. Song~\textit{et al.}~\cite{song2021through} developed a method to reconstruct 3D pose of human subjects who are hidden. Combining a UWB MIMO radar and a 3D pose reconstruction model based on camera images to generate labels, they successfully predicted the 3D pose of hidden human targets. Similarly, Zhao~\textit{et al.}~\cite{zhao2018through} estimated wall-obstructed humans using WiFi signals. Using RF and RGB data, they have built a teacher-student network and performed cross-modal supervision for accurate estimation of 2D human pose.
On the other hand, cameras can support radio devices in cases that reduce the performance of the latter, such as multi-users scenarios~\cite{Xie2021} and body movement~\cite{gu2013hybrid}, predicting the optimal beam such that the received signal power is maximized~\cite{Charan2022}.
The combination is beneficial since these two modalities can complement each other.

From camera-augmented radar to radar-guided camera approaches, researchers are demonstrating the viability of cross modal collaborative systems in addressing inherent modality limitations. However, challenges persist in optimizing these architectures, especially in ensuring real-time modality synchronization and seamless data fusion. Future research should explore dynamic modality collaboration mechanisms~\cite{Li2022_Self_Supervised}, that can adapt to unsynchronized data streams, and the investigation of adaptive modality weighting based on environmental contexts.

\section{Discussion}
In this section, we discuss shortcomings of existing multimodal datasets. Then, we address recent trends of machine learning methods and issues of multimodal data.

\subsection{Public datasets of visual and non-visual modalities}

We select several existing datasets that include visual (e.g. RGB videos and depth videos) and non-visual modalities (e.g., RF signals and inertial data).
These datasets are described succintly in Table~\ref{tab:datasets}.
Other surveys have been considered different combinations of modalities, such as RGB and depth data~\cite{Wang2023}.
\begin{table*}
\caption{Multimodal datasets and state-of-the-art results}
\label{tab:datasets}
\begin{tabular}{|p{3cm}|p{6cm}|p{8cm}|}
\hline
\textbf{Datasets}                           & \textbf{Modalities}    & \textbf{Description} \\ \hline
Berkeley MHAD~\cite{Ofli2013} & 12 RGB cameras, two Microsoft kinect cameras, six wearable acceleration sensors, and four microphones & 11 actions performed by 12 subjects (seven males and five females) \\ \hline
UTD-MHAD~\cite{Chen2015} & Kinect camera and one wearable inertial sensor & Eight subjects (four males and four females) performed 27 actions \\ \hline
UR Fall Detection~\cite{Bogdan2014} & Kinect cameras on the sideview and one inertial measurement unit was worn near the pelvis & Five persons, ADLs (walking, sitting, crouching down, leaning down/picking up objects from the floor, and lying on a settee),  three types of falls (forward, backward, and lateral) \\ \hline
CZU-MHAD~\cite{Chao2022_dataset}  & RGBD camera (Kinect), skeleton data, 10 wearable sensors & Five subjects, No description on annotation \\ \hline
Opportunity++~\cite{Ciliberto2021} & Sideview and topview cameras, inertial motion capture system, an UWB localization system, motion and switch sensors in objects & Early-morning routine of four subjects: preparing a coffee and sandwich, then having the breakfast, and finally cleaning the kitchen by putting food and utensils in place. \\ \hline



OPERAnet~\cite{bocus2022operanet} & radio frequency and vision-based sensors & 8 hours, 2 rooms, 6 participants, 6 activities, namely, sitting down on a chair, standing from sit, lying down on the ground, standing from the floor, walking and body rotating, UWB-anchor-based positioning, activities: labels captured when activities were performed (following the script or manual annotation) \\ \hline


Alkhateeb~\cite{Alkhateeb2022dataset} & radar, mmWave radio, camera, 3D LiDAR and gps receiver & The dataset is a collection of scenarios with each scenario containing multi-modal sensing and communication data. The collected scenarios are for applications such as beam prediction, user identification, object localization, object detection, and blockage identification \\ \hline 

Chen~\textit{et al.}~\cite{Chen2022} & 4D imaging radar, RGBD camera & 20 volunteers in 6 different environments. The  mmWave radar can achieve better body reconstruction accuracy than the RGB camera but worse than the depth camera \\ \hline

Topham~\textit{et al.}~\cite{Topham2023} & Two digital cameras and a wearable digital goniometer & Casual walks from 64 participants, in both indoor and outdoor real-world environments. Annotated with Human Pose Estimation using OpenPose 2D HPE23 system (75 anatomical keypoints); then, all authors manually performed a visual check. \\ \hline


\end{tabular}
\end{table*}

The datasets presented exhibit a consistent trend towards integrating both visual modalities, such as RGB and depth videos, and non-visual modalities like RF signals, inertial data, and wearables. A common pattern is the use of Kinect cameras, which offer depth information, coupled with wearables or other non-visual sensors. Subjects across datasets typically perform a range of everyday activities, ensuring diverse motion patterns for robust model training. However, the number of subjects, actions, and the specificity of annotations vary. Notably absent is the integration of newer sensing technologies and modalities not yet common in current datasets. For future datasets, there is a potential emphasis on more complex, real-world scenarios, a broader array of sensing technologies, and more detailed, possibly automated, annotation methodologies.

\subsubsection{Synchronization}

Synchronizing multimodal data streams is a cornerstone task in the successful combination and subsequent analysis of heterogeneous sensor data. Misalignment or asynchrony can introduce significant errors and adversely affect downstream processing and insights. We investigate methods to synchronize and align signals captured by different sensors at various sampling rates. Capturing multiple modalities on the same machine ensures shared timestamps, providing a direct way to reduce asynchrony~\cite{Chen2022}. Another prevalent strategy is clock configuration to a unified time server~\cite{bocus2022operanet}\cite{Chen2022}. For instance, Bocus\textit{et al.}\cite{bocus2022operanet} employed a local Network Time Protocol (NTP) server to achieve synchronization across modalities. In practical scenarios like activity recognition, actions that create distinguishable events in the data can serve as markers for the onset and conclusion of data capture sessions, such as a jump. Manual checks can identify misalignments between modalities, which are then adjusted accordingly \cite{Chen2022}.

The effective synchronization of multimodal data is imperative for meaningful analysis and application. Future efforts may focus on more automated and robust methods that can handle a wide variety of sensors, environmental conditions, and real-world scenarios.

\subsubsection{Annotation}
The process of annotation, a fundamental step in curating and preparing data for various tasks, especially in machine learning and computer vision, often hinges on human intervention. This human-centric nature can inadvertently introduce errors. Addressing such inaccuracies becomes paramount to ensuring reliable and effective datasets. Engaging multiple annotators to review the same dataset can diminish the impact of individual biases and reduce the potential for mistakes.
In the context of activity labeling, Bocus~\textit{et al.}~\cite{bocus2022operanet} implemented an application to issue verbal instruction to the subjects and recorded the timestamps.
As a backup method, an observer monitored the subjects and input the start and stop moments of the activities.
One challenge when dealing with multiple data sources is to separate and associate data.
For example, a system with camera and radio devices can monitor several users simultaneously.
The views of both modalities contain multiple targets.
Signal processing and machine learning techniques are required to isolate the targets and correspond data from two modalities to the right target.
Hence, the accuracy of the annotation depends on these techniques.
Li~\textit{et al.}~\cite{li2019making} relied on AlphaPose to annotate the videos, then used them to train the WiFi-based pose estimation model.
Topham~\textit{et al.}~\cite{Topham2023} utilized the OpenPose system for pose annotatation and later manually performed a visual verification.
In vital sign measurement, the ground-truth data are obtained from a single wearable device~\cite{chian2022vitalsigns}~\cite{He2022}~\cite{Xie2021}.

While technological advancements are steadily improving annotation techniques, the fusion of automation with human oversight continues to be a vital combination. Future research could further explore advanced algorithms and frameworks to streamline this process, by providing better pre-annotations~\cite{pearce2022combined}~\cite{He2022}, or limiting the needed human intervention with active learning schemes~\cite{Sengupta2022}.

\subsubsection{Participant recruitment}
A common issue of the existing datasets is that the ratio of female and males subjects is not balanced.
For example, Berkeley MHAD~\cite{Ofli2013} contains data of seven males and five females while CZU-MHAD~\cite{Chao2022_dataset} only has data of male participants.
Future datasets should focus on inclusiveness, ensuring representativeness and fairness of all possible user groups~\cite{Bragazzi2022}~\cite{Vilesov2022}.

\subsection{Multimodal architectures based on transformers}
\label{sec:transformers_architectures}

Transformer models~\cite{Vaswani2017} have the capability of representing diverse relations between inputs from one modality or multiple modalities (e.g., text, images, audio, and radio signals)~\cite{Xu2023_Transformers}.
We can define a cross-modal transformer of input sequences (tokens) from two modalities: images $\mathbf{X}_1$ and radio signals $\mathbf{X}_2$.
We add positional encodings (PE) to the embeddings to capture the order of tokens in each sequence:
$\mathbf{Z}_1 = PE(\mathbf{X}_1)$ and $\mathbf{Z}_2 = PE(\mathbf{X}_2)$.
Within each modality, we apply a self-attention (SA) mechanism to capture relationships between tokens:
$SA(\mathbf{Z}) = SA(\mathbf{Q}, \mathbf{K}, \mathbf{V})$, where $\mathbf{Q} = \mathbf{Z}\mathbf{W}^Q$, $\mathbf{K} = \mathbf{Z}\mathbf{W}^K$, and $\mathbf{V} = \mathbf{Z}\mathbf{W}^V$; $\mathbf{W}^Q$, $\mathbf{W}^K$, and $\mathbf{W}^V$ are the three projection matrices.
Depending on the applications, we can implement masked self-attention mechanisms to leverage additional knowledge in the transformer models~\cite{Xu2023_Transformers}.
Multiple self-attention layers can be combined to form a multi-head self-attention (MHSA) structure.
Then, the output of the multi-head self-attention layer goes through a feed-forward network (FFN).
To allow information flow between modalities, we use a cross-modal attention mechanism:
$\mathbf{Z}_1 = MHSA(\mathbf{Q}_2, \mathbf{K}_1, \mathbf{V}_1)$ and $\mathbf{Z}_2 = MHSA(\mathbf{Q}_1, \mathbf{K}_2, \mathbf{V}_2)$.
For other cross-modal attention mechanisms, Xu~\textit{et al.}~\cite{Xu2023_Transformers} provided technical details and comparison between them.

Cross-modal transformers on multimodal data create the possibility to implement self-supervised learning based on cross-modal supervision methods~\cite{pearce2022combined}~\cite{He2022} (see Section~\ref{sec:collaboration}).
For example, in activity recognition, state-of-the-art vision-based models, which are publicly available, can provide labels (i.e. self-supervised signals) to train models based on RF signals.
In deployment, the system uses mostly radio devices for classifying users' activities to ensure efficiency and privacy~\cite{Kim2023_Transformer}.
Videos captured by the camera are only investigated in emergency cases, where visual details are required to clarify the current situation.

Comparing to activity recognition, vital signs measurement, user identification, and user localization, there have been fewer articles~\cite{long2018fusion}~\cite{xu2022learned}~\cite{lu2020see} on combining cameras and radio devices for 3D scene understanding in indoor settings (see Section~\ref{sec:3d}).
Through adapting 3D scene understanding techniques for autonomous driving in outdoor environments~\cite{Sengupta2022}~\cite{Lei2023}~\cite{Kim2023_Transformer}, we can integrate characteristics of RF signal propagation and visual data to infer the 3D representation of an environment more accurately.
For example, such task can be implemented using a spatio-contextual fusion transformer~\cite{Kim2023_Transformer}, which leveraged cross-attention layers to complement spatial and contextual information of videos and radio signals.
In addition, other modalities such as audio can be integrated to enhance the 3D scene understanding task~\cite{Yun2023}.

\subsection{Explainability and security}
An important aspect for healthcare applications is the explainability of the models. Nourani~\textit{et al.} \cite{nourani2020don} discussed various challanges in Explainable AI, and their system outputs human understandable explanations for activity recognition tasks. Uddin~\textit{et al.}~\cite{uddin2021human} performed body sensor-based activity recognition system using Neural Structured Learning (NSL), and and then uses Local Interpretable Model-Agnostic Explanations (LIME) to explain the model's decisions. Schmidt~\textit{et al.}~\cite{schmidt2021riftnext} perform scene context change detection and classification based on RF signals with expert driven neural explainability.
Multimodal explainable approaches ensure that models are not only accurate but also understandable and transparent
They become crucial for increasing trust and keeping ethical standards, which are especially important in healthcare systems.

\begin{figure}
\includegraphics[width=0.5\textwidth]{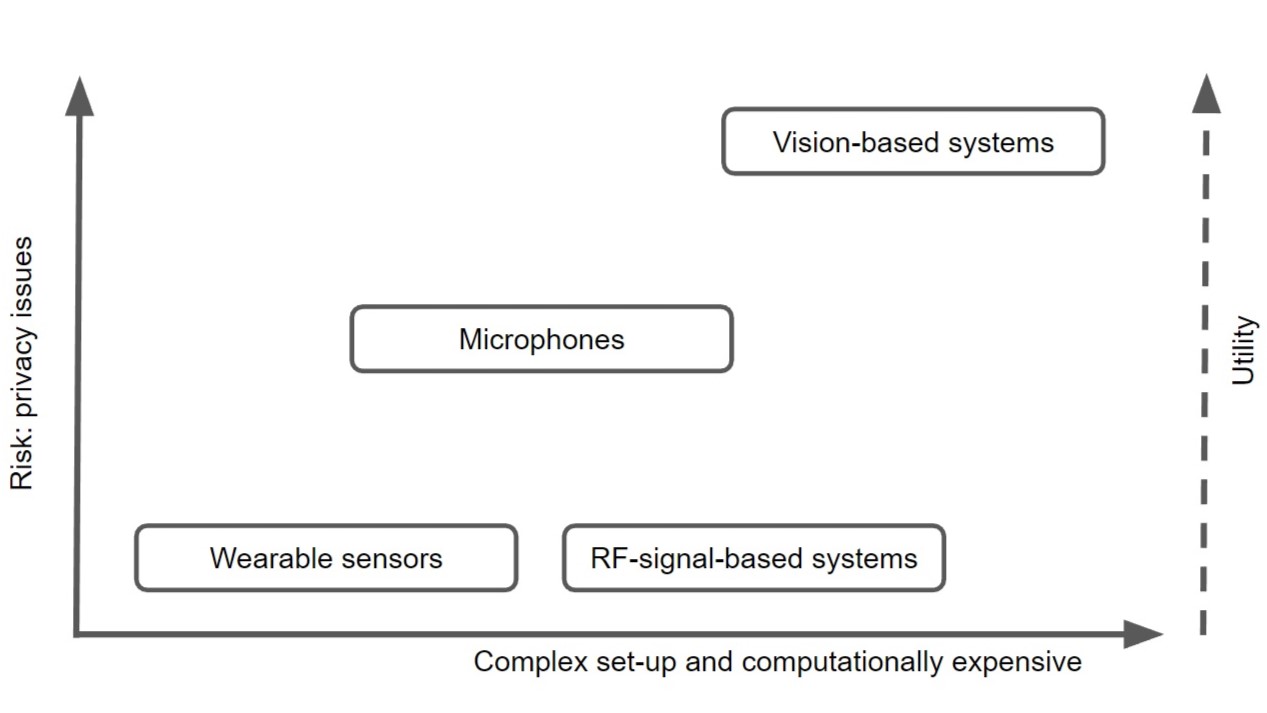}
\caption{Comparison of wearable, audio-based, vision-based, and radio-based systems}
\label{fig:multimodaltradeoff}
\end{figure}

Striking the right balance between efficient monitoring and safeguarding personal data is a critical task in every system. Figure~\ref{fig:multimodaltradeoff} provides a visual comparison of various systems, illustrating the trade-offs involved.
In the activity recognition task, this aspect of multimodal data was analyzed by Sun~\textit{et al.}~\cite{Sun2022_survey}.
We should develop a guideline to handle the security and privacy issues of multiple modalities, as well as balancing their utility and risk trade-off.

Another risk of using multiple modalities is that one compromised sensor may reveal sensitive data collected by other sensors.
This can be implemented with the same techniques to deal with partial failure of sensors in a multimodal system.
For instance, Li~\textit{et al.}~\cite{Li2022_Object} integrated LiDAR, millimeter-wave radar, and Camera for 3D object detection and tracking, in some scenarios of missing modalities.
Lee~\textit{et al.}~\cite{Lee2021} compensated corrupted data of RGB cameras, depth cameras, and force sensors in robotics.

Despite recent advances, ensuring security and maintaining privacy remain pertinent research topics~\cite{Sun2022_survey}, especially focusing on cases when the malfunction or compromise of one modality~\cite{Li2022_Object}~\cite{Lee2021} can cascade vulnerabilities throughout the whole sensing and monitoring system.

\subsection{Integrated sensing and communication}
The recent integrated sensing and communication (ISAC) paradigm~\cite{Viswanathan2020_ISAC} aims to leverage the sensing ability of communication devices to collect information of environments and users.
An ISAC system results from a coordinated or joint design of sensing and communication functionalities to achieve low-cost operation, efficient use of limited spectrum, compact-size devices, additional sensing services for wireless networks, and overall improvement of system performance.
ISAC systems offer a wide range of applications, including activity recognition~\cite{Li2022_ISAC}, localization and tracking~\cite{Adhikary2023_ISAC}, vital signs monitoring~\cite{Zhao2023_ISAC}, and environment reconstruction~\cite{Zhou2023_ISAC}.
The paradigm facilitates the design of a multimodal elderly monitoring system that unifies efficient communication and accurate sensing among diverse sensors.


\section{Conclusion}
Throughout this survey, we delved into the multifaceted world of multimodal approaches that play a pivotal role in enhancing human monitoring systems. Our primary focus was on the combination of cameras and radio devices, showcasing how they work hand-in-hand to enhance each other's strengths. Their collaboration stands out especially in elderly monitoring systems where non-contact methods are paramount for both efficiency and user comfort~\cite{Stanford2002}. We analyzed the literature in each one of the sub-tasks and components of a truly multimodal monitoring system, summarizing the recent research, enumerating the remaining challenges and devising possible lines for future research. When analyzing the available resources, we also pointed out the shortcomings in existing datasets, underscoring the need for more comprehensive and versatile data collections. In addition, we discussed potential topics of interest that can be seen throughout all technical implementations. From our analysis, it can be seen that working with multimodal data presents its own set of challenges, but the potential benefits for improving human monitoring systems are well worth the effort.

\section*{ACKNOWLEDGEMENTS}
This research has been supported by the Academy of Finland 6G Flagship program under Grant 346208 and PROFI5 HiDyn program under Grant 32629, and by the InSecTT project under the European ECSEL Joint Undertaking (JU) program grant No 876038.

\bibliographystyle{IEEEtran}
\bibliography{application1_human_activity_references, application2_vital_sign_references, application3_object_localization_references, application4_depth_3D_references, application5_identification_references, other_multimodal}

\begin{thebibliography}{100}
\providecommand{\url}[1]{#1}
\csname url@samestyle\endcsname
\providecommand{\newblock}{\relax}
\providecommand{\bibinfo}[2]{#2}
\providecommand{\BIBentrySTDinterwordspacing}{\spaceskip=0pt\relax}
\providecommand{\BIBentryALTinterwordstretchfactor}{4}
\providecommand{\BIBentryALTinterwordspacing}{\spaceskip=\fontdimen2\font plus
\BIBentryALTinterwordstretchfactor\fontdimen3\font minus
  \fontdimen4\font\relax}
\providecommand{\BIBforeignlanguage}[2]{{%
\expandafter\ifx\csname l@#1\endcsname\relax
\typeout{** WARNING: IEEEtran.bst: No hyphenation pattern has been}%
\typeout{** loaded for the language `#1'. Using the pattern for}%
\typeout{** the default language instead.}%
\else
\language=\csname l@#1\endcsname
\fi
#2}}
\providecommand{\BIBdecl}{\relax}
\BIBdecl

\bibitem{world2023progress}
{\relax World Health Organization}, ``{Progress report on the United Nations
  decade of healthy ageing, 2021-2023},'' in \emph{UN Decade of Healthy
  Ageing}, 2023.

\bibitem{Gokalp2013}
H.~Gokalp and M.~Clarke, ``Monitoring activities of daily living of the elderly
  and the potential for its use in telecare and telehealth: a review,''
  \emph{TELEMEDICINE and e-HEALTH}, 2013.

\bibitem{Aggarwal2014}
J.~K. Aggarwal and L.~Xia, ``Human activity recognition from 3d data: A
  review,'' \emph{Pattern Recognition Letters}, 2014.

\bibitem{Fernandes2022}
J.~M. Fernandes, J.~S. Silva, A.~Rodrigues, and F.~Boavida, ``A survey of
  approaches to unobtrusive sensing of humans,'' \emph{ACM Computing Surveys},
  2022.

\bibitem{wang2019wearablesensors}
J.~Wang, Y.~Chen, S.~Hao, X.~Peng, and L.~Hu, ``Deep learning for sensor-based
  activity recognition: A survey,'' \emph{Pattern recognition letters}, vol.
  119, pp. 3--11, 2019.

\bibitem{Haque2020}
A.~Haque, A.~Milstein, and L.~Fei-Fei, ``Illuminating the dark spaces of
  healthcare with ambient intelligence,'' \emph{Nature}, 2020.

\bibitem{Li2021_Crowd}
X.~Li, Q.~Yu, B.~Alzahrani, A.~Barnawi, A.~Alhindi, D.~Alghazzawi, and Y.~Miao,
  ``Data fusion for intelligent crowd monitoring and management systems: A
  survey,'' \emph{IEEE Access}, 2021.

\bibitem{shokouhmand2022camera}
A.~Shokouhmand, S.~Eckstrom, B.~Gholami, and N.~Tavassolian, ``Camera-augmented
  non-contact vital sign monitoring in real time,'' \emph{IEEE Sensors
  Journal}, 2022.

\bibitem{Qiu2022detection}
C.~Qiu, D.~Zhang, Y.~Hu, H.~Li, Q.~Sun, and Y.~Chen, ``Radio-assisted human
  detection,'' \emph{IEEE Transactions on Multimedia}, 2022.

\bibitem{Plothner2019}
M.~Pl{\"o}thner, K.~Schmidt, L.~de~Jong, J.~Zeidler, and K.~Damm, ``Needs and
  preferences of informal caregivers regarding outpatient care for the elderly:
  a systematic literature review.'' \emph{BMC Geriatrics}, 2019.

\bibitem{Stanford2002}
V.~Stanford, ``Using pervasive computing to deliver elder care,'' \emph{IEEE
  Pervasive computing}, 2002.

\bibitem{Page2021}
M.~J. Page, J.~E. McKenzie, P.~M. Bossuyt, I.~Boutron, T.~C. Hoffmann, C.~D.
  Mulrow, L.~Shamseer, J.~M. Tetzlaff, E.~A. Akl, S.~E. Brennan \emph{et~al.},
  ``The prisma 2020 statement: an updated guideline for reporting systematic
  reviews,'' \emph{International Journal of Surgery}, 2021.

\bibitem{krahnstoever2005activity}
N.~Krahnstoever, J.~Rittscher, P.~Tu, K.~Chean, and T.~Tomlinson, ``Activity
  recognition using visual tracking and rfid,'' in \emph{2005 Seventh IEEE
  Workshops on Applications of Computer Vision (WACV/MOTION'05)-Volume 1},
  vol.~1.\hskip 1em plus 0.5em minus 0.4em\relax IEEE, 2005, pp. 494--500.

\bibitem{Sun2022_survey}
Z.~Sun, Q.~Ke, H.~Rahmani, M.~Bennamoun, G.~Wang, and J.~Liu, ``Human action
  recognition from various data modalities: A review,'' \emph{IEEE Transactions
  on Pattern Analysis and Machine Intelligence}, 2022.

\bibitem{Jianwei2023RGBDReview}
J.~Li, W.~Gao, Y.~Wu, Y.~Liu, and Y.~Shen, ``High-quality indoor scene 3d
  reconstruction with rgb-d cameras: A brief review,'' \emph{Computational
  Visual Media}, vol.~8, pp. 1--25, 03 2022.

\bibitem{yousefi2017survey}
S.~Yousefi, H.~Narui, S.~Dayal, S.~Ermon, and S.~Valaee, ``A survey on behavior
  recognition using wifi channel state information,'' \emph{IEEE Communications
  Magazine}, vol.~55, no.~10, pp. 98--104, 2017.

\bibitem{Ciliberto2021}
M.~Ciliberto, V.~Fortes~Rey, A.~Calatroni, P.~Lukowicz, and D.~Roggen,
  ``Opportunity++: A multimodal dataset for video-and wearable, object and
  ambient sensors-based human activity recognition,'' \emph{Frontiers in
  Computer Science}, 2021.

\bibitem{Morita2023}
P.~P. Morita, K.~S. Sahu, and A.~Oetomo, ``Health monitoring using smart home
  technologies: Scoping review,'' \emph{JMIR mHealth and uHealth}, 2023.

\bibitem{yuan2022survey}
M.~Yuan, S.~Wei, J.~Zhao, and M.~Sun, ``A systematic survey on human behavior
  recognition methods,'' \emph{SN Computer Science}, vol.~3, no.~1, pp. 1--25,
  2022.

\bibitem{Wang2023}
C.~Wang and J.~Yan, ``A comprehensive survey of rgb-based and skeleton-based
  human action recognition,'' \emph{IEEE Access}, 2023.

\bibitem{wang2015review}
S.~Wang and G.~Zhou, ``A review on radio based activity recognition,''
  \emph{Digital Communications and Networks}, 2015.

\bibitem{Soto2022}
J.~C. Soto, I.~Galdino, E.~Caballero, V.~Ferreira, D.~Muchaluat-Saade, and
  C.~Albuquerque, ``A survey on vital signs monitoring based on wi-fi csi
  data,'' \emph{Computer Communications}, 2022.

\bibitem{tang2022survey}
X.~Tang, Z.~Zhang, and Y.~Qin, ``On-road object detection and tracking based on
  radar and vision fusion: A review,'' \emph{IEEE Intelligent Transportation
  Systems Magazine}, 2022.

\bibitem{Baltruvsaitis2018Survey}
T.~Baltru{\v{s}}aitis, C.~Ahuja, and L.-P. Morency, ``Multimodal machine
  learning: A survey and taxonomy,'' \emph{IEEE Transactions on Pattern
  Analysis and Machine Intelligence}, 2018.

\bibitem{Sleeman2022}
W.~C. Sleeman, R.~Kapoor, and P.~Ghosh, ``Multimodal classification: Current
  landscape, taxonomy and future directions,'' \emph{ACM Computing Surveys},
  2022.

\bibitem{Fan2020}
L.~Fan, T.~Li, Y.~Yuan, and D.~Katabi, ``In-home daily-life captioning using
  radio signals,'' in \emph{European Conference on Computer Vision}, 2020.

\bibitem{zhao2018through}
M.~Zhao, T.~Li, M.~Abu~Alsheikh, Y.~Tian, H.~Zhao, A.~Torralba, and D.~Katabi,
  ``Through-wall human pose estimation using radio signals,'' in
  \emph{Proceedings of the IEEE Conference on Computer Vision and Pattern
  Recognition}, 2018, pp. 7356--7365.

\bibitem{Zhao2021}
A.~Zhao, J.~Li, J.~Dong, L.~Qi, Q.~Zhang, N.~Li, X.~Wang, and H.~Zhou,
  ``Multimodal gait recognition for neurodegenerative diseases,'' \emph{IEEE
  transactions on cybernetics}, 2021.

\bibitem{Shao2022}
W.~Shao, Z.~You, L.~Liang, X.~Hu, C.~Li, W.~Wang, and B.~Hu, ``A multi-modal
  gait analysis-based detection system of the risk of depression,'' \emph{IEEE
  Journal of Biomedical and Health Informatics}, 2022.

\bibitem{zou2019wifi}
H.~Zou, J.~Yang, H.~Prasanna~Das, H.~Liu, Y.~Zhou, and C.~J. Spanos, ``Wifi and
  vision multimodal learning for accurate and robust device-free human activity
  recognition,'' in \emph{Proceedings of the IEEE/CVF conference on computer
  vision and pattern recognition workshops}, 2019, pp. 0--0.

\bibitem{ardianto2018infrared}
S.~Ardianto and H.-M. Hang, ``Multi-view and multi-modal action recognition
  with learned fusion,'' in \emph{2018 Asia-Pacific Signal and Information
  Processing Association Annual Summit and Conference (APSIPA ASC)}.\hskip 1em
  plus 0.5em minus 0.4em\relax IEEE, 2018, pp. 1601--1604.

\bibitem{de2020infrared}
A.~M. De~Boissiere and R.~Noumeir, ``Infrared and 3d skeleton feature fusion
  for rgb-d action recognition,'' \emph{IEEE Access}, vol.~8, pp.
  168\,297--168\,308, 2020.

\bibitem{memmesheimer2020}
R.~Memmesheimer, N.~Theisen, and D.~Paulus, ``Gimme signals: Discriminative
  signal encoding for multimodal activity recognition,'' in \emph{2020 IEEE/RSJ
  International Conference on Intelligent Robots and Systems (IROS)}.\hskip 1em
  plus 0.5em minus 0.4em\relax IEEE, 2020, pp. 10\,394--10\,401.

\bibitem{Li2018}
H.~Li, A.~Shrestha, F.~Fioranelli, J.~Le~Kernec, and H.~Heidari, ``Hierarchical
  classification on multimodal sensing for human activity recogintion and fall
  detection,'' in \emph{2018 IEEE SENSORS}, 2018.

\bibitem{robertson2006general}
N.~Robertson and I.~Reid, ``A general method for human activity recognition in
  video,'' \emph{Computer Vision and Image Understanding}, vol. 104, no. 2-3,
  pp. 232--248, 2006.

\bibitem{nie2021pose2room}
Y.~Nie, A.~Dai, X.~Han, and M.~Nie{\ss}ner, ``Pose2room: Understanding 3d
  scenes from human activities,'' \emph{arXiv preprint arXiv:2112.03030}, 2021.

\bibitem{han2005human}
J.~Han and B.~Bhanu, ``Human activity recognition in thermal infrared
  imagery,'' in \emph{2005 IEEE Computer Society Conference on Computer Vision
  and Pattern Recognition (CVPR'05)-Workshops}.\hskip 1em plus 0.5em minus
  0.4em\relax IEEE, 2005, pp. 17--17.

\bibitem{li2019making}
T.~Li, L.~Fan, M.~Zhao, Y.~Liu, and D.~Katabi, ``Making the invisible visible:
  Action recognition through walls and occlusions,'' in \emph{Proceedings of
  the IEEE/CVF International Conference on Computer Vision}, 2019, pp.
  872--881.

\bibitem{bocus2022uwbdataset}
M.~J. Bocus and R.~Piechocki, ``A comprehensive ultra-wideband dataset for
  non-cooperative contextual sensing,'' \emph{Scientific Data}, vol.~9, no.~1,
  pp. 1--13, 2022.

\bibitem{bocus2022operanet}
M.~J. Bocus, W.~Li, S.~Vishwakarma, R.~Kou, C.~Tang, K.~Woodbridge,
  I.~Craddock, R.~McConville, R.~Santos-Rodriguez, K.~Chetty \emph{et~al.},
  ``Operanet, a multimodal activity recognition dataset acquired from radio
  frequency and vision-based sensors,'' \emph{Scientific data}, 2022.

\bibitem{guo2018}
L.~Guo, L.~Wang, J.~Liu, W.~Zhou, and B.~Lu, ``Huac: Human activity recognition
  using crowdsourced wifi signals and skeleton data,'' \emph{Wireless
  Communications and Mobile Computing}, vol. 2018, 2018.

\bibitem{Li2019}
T.~Li, L.~Fan, M.~Zhao, Y.~Liu, and D.~Katabi, ``Making the invisible visible:
  Action recognition through walls and occlusions,'' in \emph{Proceedings of
  the IEEE/CVF International Conference on Computer Vision}, 2019, pp.
  872--881.

\bibitem{koupai2022activity}
A.~K. Koupai, M.~J. Bocus, R.~Santos-Rodriguez, R.~J. Piechocki, and
  R.~McConville, ``Self-supervised multimodal fusion transformer for passive
  activity recognition,'' \emph{arXiv preprint arXiv:2209.03765}, 2022.

\bibitem{Zhang2021}
Q.~Zhang and Y.~Li, ``Indoor positioning method based on infrared vision and
  uwb fusion,'' in \emph{Journal of Physics: Conference Series}, 2021.

\bibitem{Bragazzi2022}
N.~L. Bragazzi, R.~Khamisy-Farah, and M.~Converti, ``Ensuring equitable,
  inclusive and meaningful gender identity-and sexual orientation-related data
  collection in the healthcare sector: insights from a critical, pragmatic
  systematic review of the literature,'' \emph{International Review of
  Psychiatry}, 2022.

\bibitem{He2022}
S.~He, Z.~Han, C.~Iglesias, V.~Mehta, and M.~Bolic, ``A real-time respiration
  monitoring and classification system using a depth camera and radars,''
  \emph{Frontiers in Physiology}, 2022.

\bibitem{ren2017comparison}
L.~Ren, L.~Kong, F.~Foroughian, H.~Wang, P.~Theilmann, and A.~E. Fathy,
  ``Comparison study of noncontact vital signs detection using a doppler
  stepped-frequency continuous-wave radar and camera-based imaging
  photoplethysmography,'' \emph{IEEE Transactions on Microwave Theory and
  Techniques}, vol.~65, no.~9, pp. 3519--3529, 2017.

\bibitem{yang2021remote}
X.~Yang, Z.~Zhang, X.~Li, Y.~Zheng, and Y.~Shen, ``Remote radar-camera vital
  sign monitoring system using a graph-based extraction algorithm,'' in
  \emph{2021 46th International Conference on Infrared, Millimeter and
  Terahertz Waves (IRMMW-THz)}.\hskip 1em plus 0.5em minus 0.4em\relax IEEE,
  2021, pp. 1--2.

\bibitem{Xie2021}
Z.~Xie, B.~Zhou, X.~Cheng, E.~Schoenfeld, and F.~Ye, ``Vitalhub: Robust,
  non-touch multi-user vital signs monitoring using depth camera-aided uwb,''
  in \emph{IEEE International Conference on Healthcare Informatics}, 2021.

\bibitem{Yang2020}
C.~Yang, B.~Bruce, X.~Liu, B.~Gholami, and N.~Tavassolian, ``A hybrid
  radar-camera respiratory monitoring system based on an impulse-radio
  ultrawideband radar,'' in \emph{Annual International Conference of the IEEE
  Engineering in Medicine \& Biology Society}, 2020.

\bibitem{chian2022vitalsigns}
D.-M. Chian, C.-K. Wen, C.-J. Wang, M.-H. Hsu, and F.-K. Wang, ``Vital signs
  identification system with doppler radars and thermal camera,'' \emph{IEEE
  Transactions on Biomedical Circuits and Systems}, vol.~16, no.~1, pp.
  153--167, 2022.

\bibitem{Vilesov2022}
A.~Vilesov, P.~Chari, A.~Armouti, A.~B. Harish, K.~Kulkarni, A.~Deoghare,
  L.~Jalilian, and A.~Kadambi, ``Blending camera and 77 ghz radar sensing for
  equitable, robust plethysmography,'' \emph{ACM Transactions on Graphics},
  2022.

\bibitem{soto2022_wifivitalsign_survey}
J.~C. Soto, I.~Galdino, E.~Caballero, V.~Ferreira, D.~Muchaluat-Saade, and
  C.~Albuquerque, ``A survey on vital signs monitoring based on wi-fi csi
  data,'' \emph{Computer Communications}, vol. 195, pp. 99--110, 2022.

\bibitem{selvaraju2022_cameravitalsign_survey}
V.~Selvaraju, N.~Spicher, J.~Wang, N.~Ganapathy, J.~M. Warnecke, S.~Leonhardt,
  R.~Swaminathan, and T.~M. Deserno, ``Continuous monitoring of vital signs
  using cameras: a systematic review,'' \emph{Sensors}, vol.~22, no.~11, p.
  4097, 2022.

\bibitem{zhang2022rf}
L.~Zhang, C.~Fu, C.~Li, and H.~Hong, ``Rf and camera-based vital signs
  monitoring applications,'' in \emph{Contactless Vital Signs
  Monitoring}.\hskip 1em plus 0.5em minus 0.4em\relax Elsevier, 2022, pp.
  303--326.

\bibitem{rong2022new}
Y.~Rong, P.~C. Theofanopoulos, G.~C. Trichopoulos, and D.~W. Bliss, ``A new
  principle of pulse detection based on terahertz wave plethysmography,''
  \emph{Scientific reports}, vol.~12, no.~1, pp. 1--15, 2022.

\bibitem{cardillo2022vitalsign}
E.~Cardillo, C.~Li, and A.~Caddemi, ``Vital sign detection and radar
  self-motion cancellation through clutter identification,'' \emph{IEEE
  Transactions on Microwave Theory and Techniques}, vol.~69, no.~3, pp.
  1932--1942, 2021.

\bibitem{dai2022enhancement}
T.~K.~V. Dai, K.~Oleksak, T.~Kvelashvili, F.~Foroughian, C.~Bauder,
  P.~Theilmann, A.~E. Fathy, and O.~Kilic, ``Enhancement of remote vital sign
  monitoring detection accuracy using multiple-input multiple-output 77 ghz
  fmcw radar,'' \emph{IEEE Journal of Electromagnetics, RF and Microwaves in
  Medicine and Biology}, vol.~6, no.~1, pp. 111--122, 2022.

\bibitem{peng2021noncontact}
K.-C. Peng, M.-C. Sung, F.-K. Wang, and T.-S. Horng, ``Noncontact vital sign
  sensing under nonperiodic body movement using a novel frequency-locked-loop
  radar,'' \emph{IEEE Transactions on Microwave Theory and Techniques},
  vol.~69, no.~11, pp. 4762--4773, 2021.

\bibitem{wang2021multi}
Y.~Wang, Y.~Shui, X.~Yang, Z.~Li, and W.~Wang, ``Multi-target vital signs
  detection using frequency-modulated continuous wave radar,'' \emph{EURASIP
  Journal on Advances in Signal Processing}, vol. 2021, no.~1, pp. 1--19, 2021.

\bibitem{feng2021multitarget}
C.~Feng, X.~Jiang, M.-G. Jeong, H.~Hong, C.-H. Fu, X.~Yang, E.~Wang, X.~Zhu,
  and X.~Liu, ``Multitarget vital signs measurement with chest motion imaging
  based on mimo radar,'' \emph{IEEE Transactions on Microwave Theory and
  Techniques}, vol.~69, no.~11, pp. 4735--4747, 2021.

\bibitem{Shi2023}
Y.~Shi, L.~Du, X.~Chen, X.~Liao, Z.~Yu, Z.~Li, C.~Wang, and S.~Xue, ``Robust
  gait recognition based on deep cnns with camera and radar sensor fusion,''
  \emph{IEEE Internet of Things Journal}, 2023.

\bibitem{li2016id}
H.~Li, P.~Zhang, S.~Al~Moubayed, S.~N. Patel, and A.~P. Sample, ``Id-match: A
  hybrid computer vision and rfid system for recognizing individuals in
  groups,'' in \emph{CHI Conference on Human Factors in Computing Systems},
  2016.

\bibitem{Chen2022_RFCam}
H.~Chen, S.~Munir, and S.~Lin, ``Rfcam: Uncertainty-aware fusion of camera and
  wi-fi for real-time human identification with mobile devices,''
  \emph{Proceedings of the ACM on Interactive, Mobile, Wearable and Ubiquitous
  Technologies}, 2022.

\bibitem{Cao2022}
D.~Cao, R.~Liu, H.~Li, S.~Wang, W.~Jiang, and C.~X. Lu, ``Cross vision-rf gait
  re-identification with low-cost rgb-d cameras and mmwave radars,''
  \emph{Proceedings of the ACM on Interactive, Mobile, Wearable and Ubiquitous
  Technologies}, 2022.

\bibitem{Liu2022}
H.~Liu, A.~Alali, M.~Ibrahim, B.~B. Cao, N.~Meegan, H.~Li, M.~Gruteser,
  S.~Jain, K.~Dana, A.~Ashok, B.~Cheng, and H.~Lu, ``Vi-fi: Associating moving
  subjects across vision and wireless sensors,'' in \emph{ACM/IEEE
  International Conference on Information Processing in Sensor Networks}, 2022.

\bibitem{Deng2022_Gait}
L.~Deng, J.~Yang, S.~Yuan, H.~Zou, C.~X. Lu, and L.~Xie, ``Gaitfi: Robust
  device-free human identification via wifi and vision multimodal learning,''
  \emph{IEEE Internet of Things Journal}, 2022.

\bibitem{Fang2020}
S.~Fang, T.~Islam, S.~Munir, and S.~Nirjon, ``Eyefi: Fast human identification
  through vision and wifi-based trajectory matching,'' in \emph{International
  Conference on Distributed Computing in Sensor Systems}, 2020.

\bibitem{Luchetti2021}
A.~Luchetti, A.~Carollo, L.~Santoro, M.~Nardello, D.~Brunelli, and P.~Bosetti,
  ``Human identification and tracking using ultra-wideband-vision data fusion
  in unstructured environments,'' \emph{Acta IMEKO e-Journal of the
  International Measurement Confederation}, 2021.

\bibitem{Wan2018}
C.~Wan, L.~Wang, and V.~V. Phoha, ``A survey on gait recognition,'' \emph{ACM
  Computing Surveys}, 2018.

\bibitem{Nambiar2019}
A.~Nambiar, A.~Bernardino, and J.~C. Nascimento, ``Gait-based person
  re-identification: A survey,'' \emph{ACM Computing Surveys}, 2019.

\bibitem{Moghaddam2023}
A.~Sepas-Moghaddam and A.~Etemad, ``Deep gait recognition: A survey,''
  \emph{IEEE Transactions on Pattern Analysis and Machine Intelligence}, 2023.

\bibitem{suwannaphong2022radio}
T.~Suwannaphong, R.~McConville, and I.~Craddock, ``Radio signal strength
  indication augmentation for one-shot learning in indoor localisation,'' in
  \emph{Proceedings of the 1st ACM Workshop on Smart Wearable Systems and
  Applications}, 2022, pp. 7--12.

\bibitem{Jiang2022}
W.~Jiang, F.~Li, L.~Mei, R.~Liu, and S.~Wang, ``Visble: Vision-enhanced ble
  device tracking,'' in \emph{IEEE International Conference on Sensing,
  Communication, and Networking}, 2022.

\bibitem{Ishihara2018}
T.~Ishihara, K.~M. Kitani, C.~Asakawa, and M.~Hirose, ``Deep radio-visual
  localization,'' in \emph{IEEE Winter Conference on Applications of Computer
  Vision (WACV)}, 2018.

\bibitem{Deng2022}
K.~Deng, D.~Zhao, Q.~Han, S.~Wang, Z.~Zhang, A.~Zhou, and H.~Ma, ``Geryon: Edge
  assisted real-time and robust object detection on drones via mmwave radar and
  camera fusion,'' \emph{Proceedings of the ACM on Interactive, Mobile,
  Wearable and Ubiquitous Technologies}, 2022.

\bibitem{Li2022_Object}
Y.~Li, J.~Deng, Y.~Zhang, J.~Ji, H.~Li, and Y.~Zhang, ``{EZFusion}: A close
  look at the integration of lidar, millimeter-wave radar, and camera for
  accurate 3d object detection and tracking,'' \emph{IEEE Robotics and
  Automation Letters}, 2022.

\bibitem{Li2022_Pedestrian}
H.~Li, R.~Liu, S.~Wang, W.~Jiang, and C.~X. Lu, ``Pedestrian liveness detection
  based on mmwave radar and camera fusion,'' in \emph{IEEE International
  Conference on Sensing, Communication, and Networking}, 2022.

\bibitem{Papaioannou2014}
S.~Papaioannou, H.~Wen, A.~Markham, and N.~Trigoni, ``Fusion of radio and
  camera sensor data for accurate indoor positioning,'' in \emph{IEEE
  International Conference on Mobile Ad Hoc and Sensor Systems}, 2014.

\bibitem{Zhao2020}
Y.~Zhao, J.~Xu, J.~Wu, J.~Hao, and H.~Qian, ``Enhancing camera-based multimodal
  indoor localization with device-free movement measurement using wifi,''
  \emph{IEEE Internet of Things Journal}, 2020.

\bibitem{Cai2020}
J.~Cai and H.~Cai, ``Robust hybrid approach of vision-based tracking and
  radio-based identification and localization for 3d tracking of multiple
  construction workers,'' \emph{Journal of Computing in Civil Engineering},
  2020.

\bibitem{Xu2019_IVR}
J.~Xu, H.~Chen, K.~Qian, E.~Dong, M.~Sun, C.~Wu, L.~Zhang, and Z.~Yang, ``Ivr:
  Integrated vision and radio localization with zero human effort,''
  \emph{Proceedings of the ACM on Interactive, Mobile, Wearable and Ubiquitous
  Technologies}, 2019.

\bibitem{Nguyen2021}
T.~H. Nguyen, T.-M. Nguyen, and L.~Xie, ``Range-focused fusion of
  camera-imu-uwb for accurate and drift-reduced localization,'' \emph{IEEE
  Robotics and Automation Letters}, 2021.

\bibitem{Liu2020}
F.~Liu, J.~Zhang, J.~Wang, H.~Han, and D.~Yang, ``An uwb/vision fusion scheme
  for determining pedestrians’ indoor location,'' \emph{Sensors}, 2020.

\bibitem{Varotto2021}
L.~Varotto, A.~Cenedese, and A.~Cavallaro, ``Probabilistic radio-visual active
  sensing for search and tracking,'' in \emph{European Control Conference},
  2021.

\bibitem{streubel2016fusion}
R.~Streubel and B.~Yang, ``Fusion of stereo camera and mimo-fmcw radar for
  pedestrian tracking in indoor environments,'' in \emph{2016 19th
  International Conference on Information Fusion (Fusion)}.\hskip 1em plus
  0.5em minus 0.4em\relax IEEE, 2016, pp. 565--572.

\bibitem{pearce2022combined}
A.~Pearce, J.~A. Zhang, and R.~Xu, ``A combined mmwave tracking and
  classification framework using a camera for labeling and supervised
  learning,'' \emph{Sensors}, vol.~22, no.~22, p. 8859, 2022.

\bibitem{lim2021radical}
T.-Y. Lim, S.~A. Markowitz, and M.~N. Do, ``Radical: A synchronized fmcw radar,
  depth, imu and rgb camera data dataset with low-level fmcw radar signals,''
  \emph{IEEE Journal of Selected Topics in Signal Processing}, vol.~15, no.~4,
  pp. 941--953, 2021.

\bibitem{Cai2023}
K.~Cai, Q.~Xia, P.~Li, J.~Stankovic, and C.~X. Lu, ``Robust human detection
  under visual degradation via thermal and mmwave radar fusion,'' in
  \emph{International Conference on Embedded Wireless Systems and Networks},
  2023.

\bibitem{tarkowski2016wireless}
M.~Tarkowski, K.~Bizewski, M.~Rzymowski, K.~Nyka, and L.~Kulas, ``Wireless
  multimodal localization sensor for industrial applications,'' in \emph{2016
  21st International Conference on Microwave, Radar and Wireless Communications
  (MIKON)}.\hskip 1em plus 0.5em minus 0.4em\relax IEEE, 2016, pp. 1--4.

\bibitem{woznica2014rf}
P.~Woznica, M.~Tarkowski, M.~Plotka, and L.~Kulas, ``Rf indoor positioning
  system supported by wireless computer vision sensors,'' in \emph{2014 20th
  International Conference on Microwaves, Radar and Wireless Communications
  (MIKON)}.\hskip 1em plus 0.5em minus 0.4em\relax IEEE, 2014, pp. 1--3.

\bibitem{wang2011rfid}
C.-S. Wang and L.-C. Cheng, ``Rfid \& vision based indoor positioning and
  identification system,'' in \emph{2011 IEEE 3rd international conference on
  communication software and networks}.\hskip 1em plus 0.5em minus 0.4em\relax
  IEEE, 2011, pp. 506--510.

\bibitem{Sturari2016}
M.~Sturari, D.~Liciotti, R.~Pierdicca, E.~Frontoni, A.~Mancini, M.~Contigiani,
  and P.~Zingaretti, ``Robust and affordable retail customer profiling by
  vision and radio beacon sensor fusion,'' \emph{Pattern Recognition Letters},
  2016.

\bibitem{xu2022learned}
R.~Xu, W.~Dong, A.~Sharma, and M.~Kaess, ``Learned depth estimation of 3d
  imaging radar for indoor mapping,'' in \emph{2022 IEEE/RSJ International
  Conference on Intelligent Robots and Systems (IROS)}.\hskip 1em plus 0.5em
  minus 0.4em\relax IEEE, 2022, pp. 13\,260--13\,267.

\bibitem{lu2020see}
C.~X. Lu, S.~Rosa, P.~Zhao, B.~Wang, C.~Chen, J.~A. Stankovic, N.~Trigoni, and
  A.~Markham, ``See through smoke: robust indoor mapping with low-cost mmwave
  radar,'' in \emph{Proceedings of the 18th International Conference on Mobile
  Systems, Applications, and Services}, 2020, pp. 14--27.

\bibitem{long2018fusion}
N.~Long, K.~Wang, R.~Cheng, K.~Yang, and J.~Bai, ``Fusion of millimeter wave
  radar and rgb-depth sensors for assisted navigation of the visually
  impaired,'' in \emph{Millimetre Wave and Terahertz Sensors and Technology
  XI}, vol. 10800.\hskip 1em plus 0.5em minus 0.4em\relax SPIE, 2018, pp.
  21--28.

\bibitem{Ding2023}
H.~Ding, Z.~Chen, C.~Zhao, F.~Wang, G.~Wang, W.~Xi, and J.~Zhao, ``{MI-Mesh:
  3D} human mesh construction by fusing image and millimeter wave,''
  \emph{Proceedings of the ACM on Interactive, Mobile, Wearable and Ubiquitous
  Technologies}, 2023.

\bibitem{Guo2019}
W.~Guo, J.~Wang, and S.~Wang, ``Deep multimodal representation learning: A
  survey,'' \emph{IEEE Access}, vol.~7, pp. 63\,373--63\,394, 2019.

\bibitem{sturari2016robust}
M.~Sturari, D.~Liciotti, R.~Pierdicca, E.~Frontoni, A.~Mancini, M.~Contigiani,
  and P.~Zingaretti, ``Robust and affordable retail customer profiling by
  vision and radio beacon sensor fusion,'' \emph{Pattern Recognition Letters},
  vol.~81, pp. 30--40, 2016.

\bibitem{stotko2019albedo}
P.~Stotko, M.~Weinmann, and R.~Klein, ``Albedo estimation for real-time 3d
  reconstruction using rgb-d and ir data,'' \emph{ISPRS Journal of
  Photogrammetry and Remote Sensing}, vol. 150, pp. 213--225, 2019.

\bibitem{Muaaz2020}
M.~Muaaz, A.~Chelli, A.~A. Abdelgawwad, A.~C. Mallofré, and M.~Pätzold,
  ``Wiwehar: Multimodal human activity recognition using wi-fi and wearable
  sensing modalities,'' \emph{IEEE Access}, 2020.

\bibitem{qin2020imaging}
Z.~Qin, Y.~Zhang, S.~Meng, Z.~Qin, and K.-K.~R. Choo, ``Imaging and fusing time
  series for wearable sensor-based human activity recognition,''
  \emph{Information Fusion}, vol.~53, pp. 80--87, 2020.

\bibitem{dimitrievski2019tracking}
M.~Dimitrievski, L.~Jacobs, P.~Veelaert, and W.~Philips, ``People tracking by
  cooperative fusion of radar and camera sensors,'' in \emph{IEEE Intelligent
  Transportation Systems Conference}, 2019.

\bibitem{Cui2021tracking}
F.~Cui, Y.~Song, J.~Wu, Z.~Xie, C.~Song, Z.~Xu, and K.~Ding, ``Online
  multi-target tracking for pedestrian by fusion of millimeter wave radar and
  vision,'' in \emph{IEEE Radar Conference}, 2021.

\bibitem{Cao2022_ViTag}
B.~B. Cao, A.~Alali, H.~Liu, N.~Meegan, M.~Gruteser, K.~Dana, A.~Ashok, and
  S.~Jain, ``Vitag: Online wifi fine time measurements aided vision-motion
  identity association in multi-person environments,'' in \emph{IEEE
  International Conference on Sensing, Communication, and Networking}, 2022.

\bibitem{Piechocki2023}
R.~J. Piechocki, X.~Wang, and M.~J. Bocus, ``Multimodal sensor fusion in the
  latent representation space,'' \emph{Scientific Reports}, 2023.

\bibitem{xie_humanpose_2022}
C.~Xie, D.~Zhang, Z.~Wu, C.~Yu, Y.~Hu, Q.~Sun, and Y.~Chen, ``Accurate human
  pose estimation using rf signals,'' in \emph{2022 IEEE 24th International
  Workshop on Multimedia Signal Processing (MMSP)}.\hskip 1em plus 0.5em minus
  0.4em\relax IEEE, 2022, pp. 1--6.

\bibitem{sengupta_humnpose_2020}
A.~Sengupta, F.~Jin, R.~Zhang, and S.~Cao, ``mm-pose: Real-time human skeletal
  posture estimation using mmwave radars and cnns,'' \emph{IEEE Sensors
  Journal}, vol.~20, no.~17, pp. 10\,032--10\,044, 2020.

\bibitem{jiang_humanpose_2020}
W.~Jiang, H.~Xue, C.~Miao, S.~Wang, S.~Lin, C.~Tian, S.~Murali, H.~Hu, Z.~Sun,
  and L.~Su, ``Towards 3d human pose construction using wifi,'' in
  \emph{Proceedings of the 26th Annual International Conference on Mobile
  Computing and Networking}, 2020, pp. 1--14.

\bibitem{wang2019wifiperception}
F.~Wang, S.~Zhou, S.~Panev, J.~Han, and D.~Huang, ``Person-in-wifi:
  Fine-grained person perception using wifi,'' in \emph{Proceedings of the
  IEEE/CVF International Conference on Computer Vision}, 2019, pp. 5452--5461.

\bibitem{He2020}
S.~He, V.~Mehta, and M.~Bolic, ``A joint localization assisted respiratory rate
  estimation using ir-uwb radars,'' in \emph{Annual International Conference of
  the IEEE Engineering in Medicine \& Biology Society}, 2020.

\bibitem{song2021through}
Y.~Song, T.~Jin, Y.~Dai, Y.~Song, and X.~Zhou, ``Through-wall human pose
  reconstruction via uwb mimo radar and 3d cnn,'' \emph{Remote Sensing},
  vol.~13, no.~2, p. 241, 2021.

\bibitem{gu2013hybrid}
C.~Gu, G.~Wang, Y.~Li, T.~Inoue, and C.~Li, ``A hybrid radar-camera sensing
  system with phase compensation for random body movement cancellation in
  doppler vital sign detection,'' \emph{IEEE transactions on microwave theory
  and techniques}, vol.~61, no.~12, pp. 4678--4688, 2013.

\bibitem{Charan2022}
G.~Charan, T.~Osman, A.~Hredzak, N.~Thawdar, and A.~Alkhateeb,
  ``Vision-position multi-modal beam prediction using real millimeter wave
  datasets,'' in \emph{2022 IEEE Wireless Communications and Networking
  Conference}, 2022.

\bibitem{Li2022_Self_Supervised}
D.~Li, J.~Xu, Z.~Yang, Q.~Zhang, Q.~Ma, L.~Zhang, and P.~Chen, ``Motion
  inspires notion: Self-supervised visual-lidar fusion for environment depth
  estimation,'' in \emph{Annual International Conference on Mobile Systems,
  Applications and Services}, 2022.

\bibitem{Ofli2013}
F.~Ofli, R.~Chaudhry, G.~Kurillo, R.~Vidal, and R.~Bajcsy, ``Berkeley mhad: A
  comprehensive multimodal human action database,'' in \emph{IEEE Workshop on
  Applications of Computer Vision}, 2013.

\bibitem{Chen2015}
C.~Chen, R.~Jafari, and N.~Kehtarnavaz, ``Utd-mhad: A multimodal dataset for
  human action recognition utilizing a depth camera and a wearable inertial
  sensor,'' in \emph{IEEE International Conference on Image Processing}, 2015.

\bibitem{Bogdan2014}
B.~Kwolek and M.~Kepski, ``Human fall detection on embedded platform using
  depth maps and wireless accelerometer,'' \emph{Computer Methods and Programs
  in Biomedicine}, 2014.

\bibitem{Chao2022_dataset}
X.~Chao, Z.~Hou, and Y.~Mo, ``Czu-mhad: A multimodal dataset for human action
  recognition utilizing a depth camera and 10 wearable inertial sensors,''
  \emph{IEEE Sensors Journal}, vol.~22, no.~7, pp. 7034--7042, 2022.

\bibitem{Alkhateeb2022dataset}
A.~Alkhateeb, G.~Charan, T.~Osman, A.~Hredzak, J.~Morais, U.~Demirhan, and
  N.~Srinivas, ``Deepsense 6g: A large-scale real-world multi-modal sensing and
  communication dataset,'' \emph{IEEE Communications Magazine}, 2023.

\bibitem{Chen2022}
A.~Chen, X.~Wang, S.~Zhu, Y.~Li, J.~Chen, and Q.~Ye, ``Mmbody benchmark: 3d
  body reconstruction dataset and analysis for millimeter wave radar,'' in
  \emph{ACM International Conference on Multimedia}, 2022.

\bibitem{Topham2023}
L.~K. Topham, W.~Khan, D.~Al-Jumeily, A.~Waraich, and A.~J. Hussain, ``A
  diverse and multi-modal gait dataset of indoor and outdoor walks acquired
  using multiple cameras and sensors,'' \emph{Scientific Data}, 2023.

\bibitem{Sengupta2022}
A.~Sengupta, A.~Yoshizawa, and S.~Cao, ``Automatic radar-camera dataset
  generation for sensor-fusion applications,'' \emph{IEEE Robotics and
  Automation Letters}, 2022.

\bibitem{Vaswani2017}
A.~Vaswani, N.~Shazeer, N.~Parmar, J.~Uszkoreit, L.~Jones, A.~N. Gomez,
  {\L}.~Kaiser, and I.~Polosukhin, ``Attention is all you need,''
  \emph{Advances in neural information processing systems}, 2017.

\bibitem{Xu2023_Transformers}
P.~Xu, X.~Zhu, and D.~A. Clifton, ``Multimodal learning with transformers: A
  survey,'' \emph{IEEE Transactions on Pattern Analysis and Machine
  Intelligence}, 2023.

\bibitem{Kim2023_Transformer}
Y.~Kim, S.~Kim, J.~W. Choi, and D.~Kum, ``{CRAFT}: Camera-radar {3D} object
  detection with spatio-contextual fusion transformer,'' in \emph{Proceedings
  of the AAAI Conference on Artificial Intelligence}, 2023.

\bibitem{Lei2023}
Y.~Lei, Z.~Wang, F.~Chen, G.~Wang, P.~Wang, and Y.~Yang, ``Recent advances in
  multi-modal 3d scene understanding: A comprehensive survey and evaluation,''
  \emph{arXiv preprint arXiv:2310.15676}, 2023.

\bibitem{Yun2023}
H.~Yun, J.~Na, and G.~Kim, ``Dense 2d-3d indoor prediction with sound via
  aligned cross-modal distillation,'' in \emph{IEEE/CVF International
  Conference on Computer Vision}, 2023.

\bibitem{nourani2020don}
M.~Nourani, C.~Roy, T.~Rahman, E.~D. Ragan, N.~Ruozzi, and V.~Gogate, ``Don't
  explain without verifying veracity: An evaluation of explainable ai with
  video activity recognition,'' \emph{arXiv preprint arXiv:2005.02335}, 2020.

\bibitem{uddin2021human}
M.~Z. Uddin and A.~Soylu, ``Human activity recognition using wearable sensors,
  discriminant analysis, and long short-term memory-based neural structured
  learning,'' \emph{Scientific Reports}, 2021.

\bibitem{schmidt2021riftnext}
S.~Schmidt, J.~Stankowicz, J.~Carmack, and S.~Kuzdeba, ``Riftnext: Explainable
  deep neural rf scene classification,'' in \emph{Proceedings of the 3rd ACM
  Workshop on Wireless Security and Machine Learning}, 2021, pp. 79--84.

\bibitem{Lee2021}
M.~A. Lee, M.~Tan, Y.~Zhu, and J.~Bohg, ``Detect, reject, correct: Crossmodal
  compensation of corrupted sensors,'' in \emph{IEEE International Conference
  on Robotics and Automation}, 2021.

\bibitem{Viswanathan2020_ISAC}
H.~Viswanathan and P.~E. Mogensen, ``Communications in the 6g era,'' \emph{IEEE
  Access}, 2020.

\bibitem{Li2022_ISAC}
X.~Li, Y.~Cui, J.~A. Zhang, F.~Liu, D.~Zhang, and L.~Hanzo, ``Integrated human
  activity sensing and communications,'' \emph{IEEE Communications Magazine},
  2022.

\bibitem{Adhikary2023_ISAC}
A.~Adhikary, M.~S. Munir, A.~D. Raha, Y.~Qiao, Z.~Han, and C.~S. Hong,
  ``Integrated sensing, localization, and communication in holographic
  mimo-enabled wireless network: A deep learning approach,'' \emph{IEEE
  Transactions on Network and Service Management}, 2023.

\bibitem{Zhao2023_ISAC}
Z.~Zhao, R.~Liu, and J.~Li, ``Integrated sensing and communication based breath
  monitoring using 5g network,'' in \emph{International Wireless Communications
  and Mobile Computing (IWCMC)}, 2023.

\bibitem{Zhou2023_ISAC}
Z.~Zhou, X.~Li, J.~He, X.~Bi, Y.~Chen, G.~Wang, and P.~Zhu, ``6g integrated
  sensing and communication - sensing assisted environmental reconstruction and
  communication,'' in \emph{IEEE International Conference on Acoustics, Speech
  and Signal Processing}, 2023.

\end{thebibliography}
\end{document}